\documentclass{aa}  

\usepackage{orcidlink}
\usepackage{natbib}
\bibpunct{(}{)}{;}{a}{}{,}

\usepackage{amsmath}
\usepackage{graphicx}
\usepackage{epstopdf}
\usepackage{color}
\usepackage{tabularx} 
\usepackage{soul} 
\usepackage{booktabs} 
\usepackage{longtable}
\usepackage{subcaption}
\usepackage{booktabs}
\usepackage{multirow} 
\usepackage{rotating}
\usepackage{threeparttable}
\usepackage{txfonts}

\begin{document}

   \title{Photometric variability of blue supergiants: Constraints on stellar masses and evolutionary stages}

   \author{Alberici Adam, A.\inst{1,2}
          \and
          Cidale, L. S.\inst{1,2}
          \and
          Kourniotis, M.\inst{3}
          \and
          Glatzel, W.\inst{4}
          \and
          S\'anchez Arias, J. P.\inst{3}
          \and
          Ru\'iz~D\'iaz, M.\inst{1,2}
          \and
          Christen, A. \inst{5}
          }

   \institute{Instituto de Astrofísica de La Plata, CONICET-UNLP, Paseo del Bosque s/n, B1900FWA, La Plata, Argentina.\\ 
    \email{aldialb@fcaglp.unlp.edu.ar}
    \and
    Facultad de Ciencias Astronómicas y Geofísicas, Universidad Nacional de La Plata (UNLP), Paseo del Bosque s/n, B1900FWA, La Plata, Argentina.
    \and
    Astronomical Institute of the Academy of Sciences of the Czech Republic, Fričova 298, CZ-25165 Ondřejov, Czech Republic.
    \and
    Institut für Astrophysik, Georg-August-Universität Göttingen, Friedrich-Hund-Platz 1, D-37077 Göttingen, Germany.
    \and
    Instituto de Estadística, Universidad de Valparaíso, Av. Gran Bretaña 1111, Casilla 5030, Valparaíso, Chile.}

   \date{}

 
  \abstract
   {Blue supergiants are crucial to understanding the evolution of massive stars. However, their formation, mass loss, and pulsational properties remain poorly understood. Although distinguishing between pre- and post-red supergiant (RSG) phases is challenging, stellar pulsations provide valuable constraints on their masses and evolutionary status.} 
   {This study analyzes TESS light curves of six blue supergiants with high luminosity-to-mass ratios to identify periodicities. We performed theoretical modeling of the radial unstable modes to characterize the strange modes. We aim to estimate asteroseismic masses, compare them with spectroscopic and evolutionary mass estimates, and assess the evolutionary stage of the stars.} 
   {We employed the Lomb-Scargle periodogram and the weighted wavelet Z-transform to address quasiperiodic behavior and time-varying signals. A linear nonadiabatic stability analysis was performed to test for radial perturbations, generating eigenfrequencies that enable stellar mass estimation. In addition, we characterized the stochastic low-frequency variability (SLFV).}
  {Using linear nonadiabatic pulsation models, we derive a first guess at the asteroseismic masses. 
  We find that \object{HD~74371} and \object{HD~79186} are evolving blueward following the RSG phase with masses of $7.3$~M$_{\odot}$ and $10.7$~M$_{\odot}$, respectively, while the evolutionary stage of \object{HD~115842} (with $20.1$~M$_{\odot}$) cannot be uniquely determined with the available mass estimates. We also find that \object{HD~52382} (either $3.9$~M$_{\odot}$ or $6.1$~M$_{\odot}$) and \object{HD~99953} ($3.8$~M$_{\odot}$) are likely in the post-RSG phase; however, their low masses are difficult to reconcile with the available evolutionary models. The extreme supergiant \object{HD~80077}, with $132.7$~M$_{\odot}$, shows the largest mass discrepancy in the sample, suggesting a more complex evolutionary history. The SLFV properties do not provide a clear diagnostic of the evolutionary stage. However, we find a bimodal distribution of the power-law slopes and characteristic timescales.}
  {These results highlight the potential of asteroseismology as a complementary tool for constraining the evolutionary status of B-type supergiants. However, the mass discrepancies found for some objects indicate that additional observational and theoretical efforts are required.}
 

   \maketitle

\section{Introduction}

\begin{table*}[]
\caption{Stellar parameters and spectral types taken from \citet{ref-haucke-paper}.}
\centering
\begin{tabular}{cllcccccc}
\hline
\rule{0pt}{1.2em}%
HD Number  & Name & Sp.~type & T$_{\mathrm{eff}}$ & R$_{\star}$ &  $\log(L)$ & $\log g_{\mathrm{cor}}$ & $\log(L/M)$ & $v\sin i$  \\
& & & [K] & [$R_{\odot}$] & [dex] & [dex] & [dex] & [$\mathrm{km\,s^{-1}}$] \\
\hline
\hline
\rule{0pt}{1.2em}%
$52382$ &  & B1 Ia & $21500\pm1000$ & $21\pm2.5$ &  $4.94 \pm 0.13$ & $2.48$ & $4.29$  & $55$ \\ 
\rule{0pt}{1.2em}%
$74371$ & LN Vel &  B6 Iab/b & $13700\pm500$ & $73\pm10$ & $5.23 \pm 0.14$ & $1.81$ & $4.17$ & $30$ \\ 
\rule{0pt}{1.2em}%
 $79186$ & GX Vel & B5 Ia & $15800\pm500$ & $61\pm7$ & $5.33 \pm 0.11$ & $2.02$ & $4.21$ & $40$  \\ 
\rule{0pt}{1.2em}%
$80077$ & PV Vel &  B2 Ia+e & $17700\pm1000$ & $195\pm47$ &  $6.53 \pm 0.23$  & $2.20$ & $4.23$ & $10$ \\ 
\rule{0pt}{1.2em}%
$99953$ & V808 Cen & B1/2 Iab/b & $19000\pm1000$ & $25\pm2$ &  $4.87 \pm 0.12$ & $2.33$ & $4.22$ & $50$ \\ 
\rule{0pt}{1.2em}%
$115842$ & ALS 3038 &  B0.5 Ia/ab & $25500\pm1000$ & $35\pm4$ &  $5.67 \pm 0.12$  & $2.76$ &  $4.30$ & $50$ \\ 
\hline
\end{tabular}
\tablefoot{The luminosity, $L$, and stellar mass, $M$, are in solar units.}
\label{table:observations}
\end{table*}

There are still important open questions about massive stars that profoundly impact our understanding of their future evolution. These questions involve the formation process of massive stars \citep{Tan2014}, the amount of mass lost via stellar winds \citep{Venero2024}, the interplay between mass loss and stellar pulsations \citep{ref-kraus2015, Cidale2023}, the long-standing mass-discrepancy problem \citep{Herrero1992, Crowther2006, Searle2008, Aidelman2012}, the fraction of binary mergers \citep{Langer2022}, among others. Massive stars of more than $8$~M$_\sun$ remain as early B- or O-type stars in the main sequence (MS) for most of their lifetimes. Beyond this phase, the transition to red supergiants (RSGs) occurs gradually over several $10^4$ years at an almost constant stellar luminosity while hydrogen-shell burning becomes the dominant energy source. During this evolution, stars pass through the blue supergiant (BSG) phase. In general, stars with initial masses between $20-45$~M$_{\odot}$ undergo a "blue-loop" phase  \citep{ref-ekstr2012}, evolving back toward the BSG region during core helium burning. In contrast, stars with initial masses below $20$~M$_{\odot}$ may undergo blue loops before returning to the red region of the Hertzsprung-Russell (HR) diagram \citep{Sanchez2023}. However, the mass limit separating these evolutionary paths depends on the mixing processes and mass-loss rate, two critical ingredients in stellar evolution, which are often uncertainly modeled. This makes it observationally challenging to identify whether BSGs are in a pre- or post-RSG phase since it cannot be understood from the location of the star on the HR diagram alone. Distinguishing between the two pathways might be plausible by studying the pulsation modes of BSG stars \citep{ref-saio2013}, through photometric and spectroscopic analyses \citep{ref-saio2006, ref-kraus2012, ref-kraus2015, Sanchez2023}. Specifically, this distinction relies on the fact that significant mass loss after going through the RSG stage significantly increases the luminosity-to-mass ratio ($L/M$) than during the first crossing. \cite{Saio2013} showed that radial pulsations are excited in the BSG region only in models that were previously RSGs, while, at a given effective temperature, these models also exhibit a much richer spectrum of non-radial pulsation modes than their pre-RSG counterparts. These pulsations cause either slight variations or slight changes in the broadening of line profiles \citep{ref-simondiaz2010, ref-aerts2009, ref-aerts2010, Bowman2023, deBurgos2023}. 
 
Typical oscillation modes observed in BSGs are p and g modes excited by the $\kappa$ mechanism at the Fe bump of opacity \citep[see, e.g.,][]{Pamyatnykh1999} and stochastic excitations caused by different mechanisms \citep{ref-mathis2014, ref-aerts2015}. For stars with $L/M$ (in solar units) exceeding $10^4$, the stellar envelopes are characterized by short thermal timescales and are dominated by radiation pressure \citep{Gautschy1990b, Glatzel2024}. In this regime, the standard spectrum of acoustic modes is crossed by additional modes, the so-called strange modes, that cannot be explained by classical $\kappa$ and $\epsilon$ mechanisms \citep{Gautschy1990b, Glatzel1994, yadav2016}. The adiabatic approximation, which assumes an infinite thermal timescale, is no longer valid for describing the pulsational properties of such stars, and therefore a nonadiabatic stability analysis is needed \citep{Glatzel1994, Glatzel1998, Saio2013, Glatzel2024}. These strange mode instabilities persist over the entire range of effective temperatures from red to BSGs and have been associated with variable mass-loss phenomena \citep{Glatzel1999, ref-kraus2015, ref-haucke-paper, Cidale2023, Glatzel2024}, with direct consequences for their evolutionary trajectory and final fate.

In this study, we aim to identify the oscillation modes and analyze the frequencies of selected southern BSGs as a means for constraining their properties. Nowadays, satellite data enable studies of light curves of massive stars over long time intervals with a temporal resolution of a few minutes, allowing for the detection of "micro" oscillations. A good example is the Transiting Exoplanet Survey Satellite (TESS) space mission \citep{ref-ricker2014}. The availability of high-precision TESS photometric data provides an excellent opportunity to investigate six BSGs (\object{HD~52382}, \object{HD~74371}, \object{HD~79186}, \object{HD~80077}, \object{HD~99953}, and \object{HD~115842}) selected from the sample of \citet{ref-haucke-paper}. All exhibit $L/M>10^4$, in solar units, making them promising candidates for strange-mode pulsations. By analyzing their light curves, we aim to characterize their variability and derive constraints on their physical properties while minimizing the risk of identifying spurious periods or aliases \citep{ref-frescura, ref-baluev}. We also aim to compare the observed frequencies with the predictions from linear, nonadiabatic pulsation models. This comparison allows us to estimate the asteroseismic masses of the stars, which can then be contrasted with those derived from evolutionary models or spectroscopic techniques, thereby providing insights into the mass discrepancy problem.

The paper is structured as follows: the observations are detailed in Sect.~\ref{sec:Observations}. The time-series analysis methods and the models used in the linear stability analysis are presented in Sect.~\ref{sec:3}. Our main results and discussion are presented in Sects.~\ref{sec:4} and \ref{sec:5}, respectively. Additional data and detailed results are compiled in Appendix~\ref{appexA}. Section~\ref{sec:6} summarizes our conclusions.

\section{Target selection and TESS data}\label{sec:Observations}

The objects listed in Table~\ref{table:observations} were selected from \citet{ref-haucke-paper} based on their high $\log(L/M)$. The final sample comprises stars with effective temperatures between approximately $13\,000$~K and $26\,000$~K. These stars were selected to provide approximately uniform coverage across this temperature range. According to the variability classification of these objects, four -- (\object{HD 74371}, \object{HD 80077}, \object{HD 99953}, and \object{HD 115842}) -- are $\alpha$ Cygni variables. For these stars, the variations might not have a single explanation or even be caused by pulsations. Rather, they may result from some other surface-layer instability \citep{Krticka2021, Guzik2024}. The stellar parameters of our sample, such as the effective temperature ($T_{\rm eff}$), stellar radius ($R_{\star}$), surface gravity corrected by the centrifugal acceleration ($\log\,g_{\rm cor}$), and projected rotational velocity $v \sin i$ (where $v$ is the equatorial rotational velocity), were taken from \citet{ref-haucke-paper}, who performed a quantitative spectroscopic analysis of the photospheric lines. All values are summarized in Table~\ref{table:observations}.

For the photometric analysis, we retrieved high-cadence (120\,s) TESS light curves from the Mikulski Archive for Space Telescopes (MAST) via the \textsc{Python} package \textsc{astroquery.mast} \citep{2019AJ....157...98G}. Targets were identified using their TESS Input Catalog (TIC) designation \citep[v8.2,][]{2021arXiv210804778P}, selecting the entry closest to the coordinates of each target. The time series correspond to \texttt{PDCSAP\_FLUX} values processed by the Science Processing Operations Center (SPOC) pipeline, which applies co-trending basis vector corrections to remove systematic effects. The observations span three sectors for \object{HD~52382} and \object{HD~115842}, four sectors for \object{HD~99953}, and six sectors for \object{HD~74371}, \object{HD~79186}, and \object{HD~80077}. Each sector was detrended independently by fitting a second-order polynomial to the time series. The normalized flux ($F_{\rm n}$) was then converted to differential magnitude via $\Delta m = -2.5\log(F_{\rm n})$, with uncertainties propagated from the flux errors.\\

\section{Methodology}\label{sec:3}

\subsection{Time series analysis methods}\label{freqanalysis}

For a detailed study of the light curves of massive stars, it is important to have signal analysis tools that complement each other to provide a complete analysis \citep{Alberici2023}. Therefore, in this work, we computed the Lomb-Scargle periodogram \citep[]{ref-lomb, Scargle1982}, a well-known method for analyzing stationary signals. This method decomposes unequally spaced time-series data into a sum of sine and cosine functions, aiming to minimize the difference between observed and modeled data. On the other hand, the BSGs can exhibit erratic and quasiperiodic behavior leading to strong photometric and spectroscopic variations \citep[see, e.g.,][]{ref-lefever2007, Clark2010, ref-kraus2015, Kourniotis2025}. Therefore, we also employed the weighted wavelet Z-transform (WWZ), a method proposed by \citet{Foster1996}, based on the Morlet wavelet function. Decomposing a time series into time-frequency space allows us to determine the dominant modes and how those modes vary in time \citep{Torrence1998}. The wavelet transform provides information on the time evolution of the parameters (frequency, period, amplitude, and power) used to describe periodic and quasiperiodic signals (see Fig.~\ref{fig:analysis_HD52382}). The WWZ scalogram is essential for interpreting the results, helping us distinguish between coherent and quasiperiodic frequencies. Coherent pulsation modes are expected to appear as constant frequencies in the scalogram, whereas fluctuations in power and frequency may arise from beating between close frequencies, non-sinusoidal modulations \citep{Alberici2023, ref-bravo2014}, rotational effects \citep{Li2019, Vitali2026}, the intrinsic resolution of the WWZ technique itself \citep{Foster1996}, among others. A detailed description of the implementation of the methods and the criteria applied for the selection of frequencies is provided in Appendix~\ref{appexA}.

\subsection{Stochastic low-frequency variability} \label{sec:slfv}

For massive OB-type stars, a gradual increase in power toward low frequencies is ubiquitously observed in the frequency domain, an effect known as stochastic low-frequency variability \citep[SLFV;][]{Blomme2011, Rauw2019, Naze2021, Gong2022a}. This type of variability is intrinsic to a star, but the origin of the signal is yet unknown, with several possible explanations proposed, such as inertia-gravity waves excited at the interface of the convective core \citep{Bowman2018a, Bowman2019, Thompson2024, Kourniotis2025}, stochastic waves generated by subsurface convection or granulation \cite{Blomme2011, Cantiello2021}, damped gravity waves \citep{Ma2024}, and line-driven wind instabilities \citep{Krticka2021}. 

The SLFV of the stars in our sample, with the exception of \object{HD 52382} and \object{HD 80077}, was recently characterized by \citet{Kourniotis2025} by fitting the amplitude spectrum with a $\gamma$-index power law:
\begin{equation}
   \label{eq:Slfv}
      A(f) = \frac{A_{0}}{1 + (2\pi\tau f)^\gamma} + W,
\end{equation}
where $f$ is the frequency, $A_{0}$ is the red-noise amplitude as $f\to0$, $\tau$ is the characteristic timescale of the structures producing the variability, $\gamma$ is the slope of the power-law decay, and $W$ is the white-noise level. The presence of SLFV in the amplitude spectra of these objects reflects stochastic, turbulent variability that must be accounted for in our frequency analysis. To better understand the variability of the sample and compare the characteristic parameters of the SLFV of the stars, they must be measured using the same methodology. Therefore, we calculated the SLFV parameters for \object{HD~52382} and \object{HD~80077} using the \citet{Kourniotis2025} method. We refer the readers to that paper for a detailed explanation.

\begin{figure}[t]
    \centering
    \includegraphics[width=0.9\columnwidth]{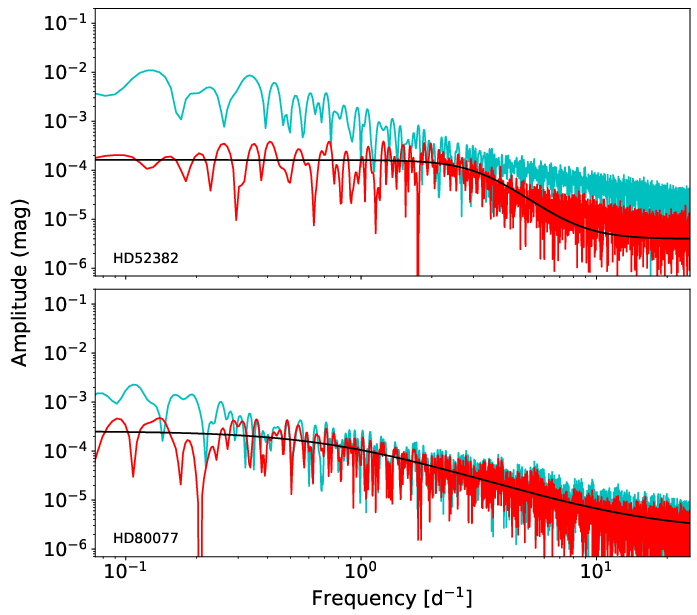}
    \caption{Frequency spectra of \object{HD~52382} (sector~$7$) and \object{HD~80077} (sectors~$35+36$). We show the periodograms at the beginning (cyan) and end (red) of the pre-whitening process following the methodology of \cite{Kourniotis2025}. The solid black line corresponds to the best-fit SLFV model adopted for these stars.} 
    \label{fig:SLFV}
\end{figure}

Our fittings are shown in Fig.~\ref{fig:SLFV} and the resulting parameters for all the stars are listed in Table~\ref{tab:periods}. The reported parameters reveal a clear bimodal structure in the power-law slope $\gamma$, which separates the sample into two physically distinct groups. The hotter, more compact supergiants (\object{HD~52382}, \object{HD~99953}, and \object{HD~115842}) exhibit steep slopes $\gamma$ and short characteristic timescales $\tau$. On the contrary, the cooler, more extended objects (\object{HD~74371}, \object{HD~79186}, and \object{HD~80077}) show flatter slopes and longer characteristic timescales. A further discussion of these results in a broader context is presented in Sect.~\ref{sec:5}.

\begin{table*}[ht]
\caption{Periods and SLFV parameters for the sample stars.}
\centering
\begin{tabular}{c c c c c c c c}
\hline
\rule{0pt}{1.2em}%
ID & Var. type & (P$_{known}$)~[d] & P$_{TESS}$~[d] & $\log(W)$ & $\log(A_0)$ & $\tau$ & $\gamma$ \\
\hline
\hline
\rule{0pt}{1.2em}%
HD 52382 & IA & - & $1.1,~2.3,~2.9$ & $-5.40$ & $-3.80$ & $0.05$ & $4.04$ \\
&    &   & $3.4,~8.7$ &&&& \\
\hline
\rule{0pt}{1.2em}%
HD 74371 & ACYG, IA & $2.9^a,~5.2^a,~7.7^a$ & $2.5,~3.1,~4.1$ & $-5.51^\dagger$ & $-3.72^\dagger$ & $0.13^\dagger$ & $1.87^\dagger$ \\
&  & $8.3^{b,f}$ & $5.4,~7.3,~8.8$ & &&& \\
\hline
\rule{0pt}{1.2em}%
HD 79186 & IA & $78.6^c$ & $3.2,~3.9,~5.0$ & $-5.30^\dagger$ & $-3.11^\dagger$ & $0.25^\dagger$ & $1.58^\dagger$ \\
&    &   & $6.1,~8.7$ &&&& \\
\hline
\rule{0pt}{1.2em}%
HD 80077 & GCAS, SDOR  & $3.1^b,~21.2^e,~41.5^g$  & $3.5,~4.0,~5.4$ & $-5.63$ & $-3.60$ & $0.20$ & $1.59$ \\
& ACYG & $55.5^g,~66.5^g,~76.0^g$  & $8.6$  \\
\hline
\rule{0pt}{1.2em}%
HD 99953 &  ACYG, IA   & $17.7^d$ & $2.5,~3.1,~4.0$ & $-5.26^\dagger$ & $-3.75^\dagger$ & $0.07^\dagger$ & $3.80^\dagger$ \\
&  & & $8.8$ & &&& \\
\hline
\rule{0pt}{1.2em}%
HD 115842 & ACYG, IA & $10.3^b,~13.4^f$ & $1.4,~2.1,~4.0$ & $-5.51^\dagger$ & $-3.58^\dagger$ & $0.05^\dagger$ & $3.58^\dagger$ \\
 &          &    & $6.0,~9.2$ & &&& \\
\bottomrule
\end{tabular}
\tablefoot{
Columns 2, 3, and 4 list the variability type, periods reported in the literature (P$_{known}$), and photometric periods (P$_{TESS}$) determined in this work, respectively. The SLFV parameters defined by Eq.~(\ref{eq:Slfv}) are listed in the last four columns. Parameters marked with a dagger ($^\dagger$) were taken from \citet{Kourniotis2025} and are shown for comparison. Those for \object{HD~52382} and \object{HD~80077} were derived in this work. More details on the detected frequencies and periods are given in Table~\ref{table:all}. The variability classification was taken from  \cite{ref-lefevre2009}: IA (poorly studied variables of early spectral type), GCAS ($\gamma$ Cassiopeiae type), ACYG ($\alpha$~Cygni type), and S Dor (S Doradus type). The periods reported in the literature were compiled from: $^a$\citep{vanGenderen1989},  $^b$\citep{ref-lefevre2009},  $^c$\citep{Underhill1984},  $^d$\citep{Sterken1977}, $^e$\citep{Knoechel1982}, $^f$\citep{ref-kohen}, and $^g$\citep{Leeuwen1998}.
}
\label{tab:periods}
\end{table*}

\subsection{The linear nonadiabatic model}

In this subsection, we present the linear nonadiabatic pulsation framework used to investigate the stability properties of stellar models. The first step in such a theoretical analysis is the construction of appropriate hydrostatic stellar models, which serve as the equilibrium configurations for the subsequent pulsation calculations \citep{Ober1983, GautschyGlatzel1990, Gautschy1990b, Kiriakidis1993}. As the chemical composition, luminosity, effective temperature, and, at least, a guess for the mass of the objects are provided by the observations, it seems reasonable to calculate hydrostatic stellar models by an envelope integration from the photosphere into the interior of the star up to some conveniently chosen cut - off - temperature, typically on the order of $10^7$ K  \citep{Yadavetal2021}. An envelope integration (solving the equations of mass conservation, hydrostatic equilibrium, and energy transport) poses an initial value problem with unique initial conditions specified at the photosphere, i.e., using the prescribed parameters: stellar luminosity, effective temperature, and mass, together with an estimate of the photospheric pressure. For the opacity, we adopted the latest OPAL tables from \citet[][]{IglesiasRogers1996}.  Convection was treated according to the standard mixing length theory \citep[see, e.g.,][]{BoehmVitense1958} using 1.5 pressure scale heights for the mixing length. With respect to the mass uncertainty for each object, a sequence of models was generated with a fixed luminosity and effective temperature (as provided by the observations), with the stellar mass being a free parameter within an interval of values comprising the observational guess (see Sect.~\ref{sec:modelsequence}). Subsequently, the envelope models were tested for linear nonadiabatic stability in response to radial perturbations. The associated perturbation equations and appropriate boundary conditions were accounted for as explained, for example, in \citet{BakerKippenhahn1962}. Convection was treated within the frozen-in-approximation \citep[see also, e.g.,][]{BakerKippenhahn1965}. 

As a result of the stability analysis, we obtained a discrete set of complex eigenfrequencies $\sigma$, whose real parts correspond to oscillation frequencies. Positive imaginary parts indicate damping; negative imaginary parts indicate growth and instability of a mode. For convenience, $\sigma$ is usually made dimensionless using the global free-fall timescale. The numerical solution of the boundary eigenvalue problem associated with the linear nonadiabatic stability analysis faces severe difficulties for stellar envelopes, which are characterized by small thermal timescales and a significant fraction of radiation pressure. Such models, to which the sequences considered here belong, exhibit strange modes together with violent instabilities with growth rates in the dynamical range. Therefore, we adopted the Riccati method \citep[see][]{GautschyGlatzel1990} for a numerical treatment of the problem, which has been proven to provide reliable and highly precise results even in highly nonadiabatic environments.

\section{Results}\label{sec:4}

\subsection{Frequency analysis}\label{sec:photometricanalysis}

We analyzed the TESS light curves of the target stars to identify their independent frequencies. Significant frequencies were extracted through an iterative pre-whitening procedure. To complement the Fourier analysis and assess the temporal stability of the detected signals, we also applied the WWZ method. A detailed presentation of the corresponding periodograms and WWZ scalograms is provided in Appendix~\ref{appexA}.

\object{HD~52382} (B1Ia) is a luminous supergiant located on the border of the constellation of Monoceros, the Unicorn. It is listed as an O-B2 runaway candidate \citep{Tetzlaff2011}. Pulsational activity was suggested based on the high observed macroturbulence \citep{ref-haucke-paper, Lefever2010}; however, no pulsation periods have previously been documented for this star.

\begin{figure}[ht!]
    \centering    
    \includegraphics[width=1.0\linewidth]{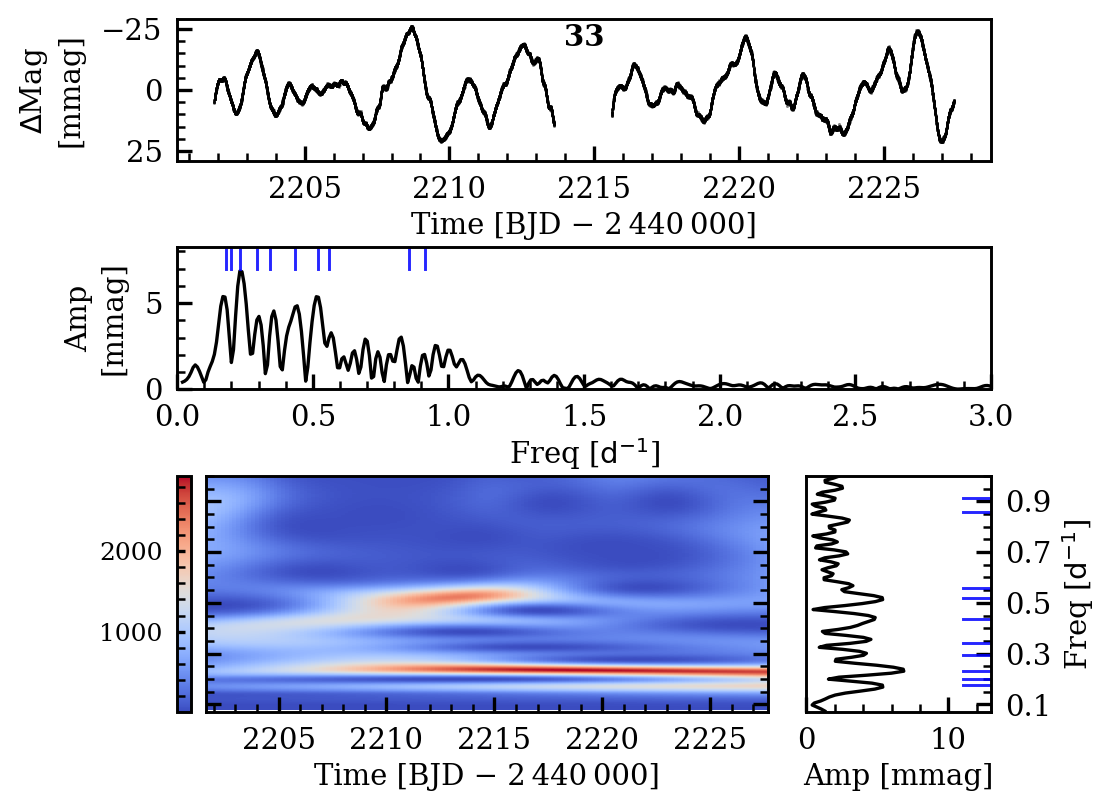}
    \caption{\textit{Top panel:} TESS light curve of HD 52382 from sector $33$. \textit{Middle panel:} Lomb–Scargle periodogram, showing the global frequency spectrum and the frequencies detected  (blue tick marks). \textit{Bottom panel:} (\textit{left}) WWZ scalogram (power as a function of frequency and time), and (\textit{right}) zoom-in on the Lomb–Scargle periodogram.}
    \label{fig:analysis_HD52382}
\end{figure}

Our analysis of TESS sectors~$7$, $33$, and $87$ reveals five independent frequencies (whose associated periods are listed in Table~\ref{tab:periods}), with most showing amplitude variations from sector-to-sector. Two ($f_{1}$ and $f_{2}$) also exhibit frequency variability (see Table~\ref{table:all}). Several harmonic and beating frequencies are additionally detected in the WWZ scalogram (Fig.~\ref{fig:analysis_HD52382}). The first independent frequency, $f_{1} = 0.115$\,d$^{-1}$~($P\approx8.7$\,days), is detected in sectors~$7$ and~$87$ but is absent in sector~$33$. The observed frequency difference between both detections remains within the Rayleigh frequency resolution and is interpreted as the same mode undergoing temporal amplitude and frequency modulation. The second frequency, $f_{2} = 0.293$\,d$^{-1}$~($P\approx3.4$\,days), shows a progressive decrease in frequency across the three sectors. In contrast, $f_{3} = 0.343$\,d$^{-1}$~($P\approx2.9$\,days), $f_{4} = 0.432$\,d$^{-1}$~($P\approx2.3$\,days), and $f_{5} = 0.912$\,d$^{-1}$~($P\approx1.1$\,days) show only marginal sector-to-sector scatter with no monotonic trend, suggesting more stable oscillations.

\object{HD~74371} (B6\, Iab/b) and \object{HD~79186} (B5\, Ia) are both located in the constellation of Vela. For \object{HD~74371}, a previous photometric study using HIPPARCOS data suggested a period of $8.291$\,days \citep{ref-lefevre2009}, while \citet{vanGenderen1989} reported variations with periods of $2.9$, $5.20$, and $7.7$\,days, although these results were inconclusive. For \object{HD~79186}, \citet{Underhill1984} reported ultraviolet light variations and estimated a rotational period of approximately $78.6$\,days, proposing that the brightness changes could be due 
to the rotation of a mottled disk.

Our analysis of the combined TESS sectors $8+9$, $35+36$, $61+62$, and $62+63$ reveals a complex low-frequency variability pattern for both stars, with several independent frequencies identified alongside numerous transient peaks, beating frequencies, and sector-dependent signals. This is also evident in the WWZ scalograms, where several structures vary in amplitude and coherence over time. Given their similar spectral classifications, location, and comparable frequency distributions, both stars appear to share a similar variability behavior.

For \object{HD~74371}, six independent frequencies were identified between $0.11$ and $0.40$\,d$^{-1}$, corresponding to periods between approximately $2.5$ and $8.8$ days, with the dominant frequencies concentrated between $0.136$ and $0.246$\,d$^{-1}$. Several frequencies display moderate sector-to-sector variations in amplitude and frequency. Notably, $f_{1}$~($P\approx8.8$\,days), $f_{2}$~($P\approx7.3$\,days), $f_{3}$~($P\approx5.4$\,days), and $f_{5}$~($P\approx3.1$\,days) aligns the periods previously reported by \citet{vanGenderen1989}. Similarly, \object{HD~79186} exhibits five independent frequencies distributed between $0.11$ and $0.31$~d$^{-1}$, corresponding to periods from $3.2$ to $8.7$ days. The frequency pattern is remarkably similar to that of \object{HD~74371}, especially for the dominant low-frequency component near $0.11$~d$^{-1}$ and the sequence of peaks around $0.20$--$0.31$~d$^{-1}$ (Figs.~\ref{fig:freqHD74371} and \ref{fig:freqHD79186}).

Both stars additionally display subtle visual micro-oscillations in their light curves over two consecutive sectors: \object{HD~74371} in sectors~$61+62$ and \object{HD~79186} in sectors~$62+63$. These oscillations fall below the adopted $\mathrm{S/N}$ cutoff and were therefore excluded from the frequency analysis. They are more likely associated with stochastic surface variability rather than coherent stellar p modes, since they do not persist across different sectors.

\begin{figure*}[ht!]
    \centering
    \includegraphics[width=0.9\linewidth]{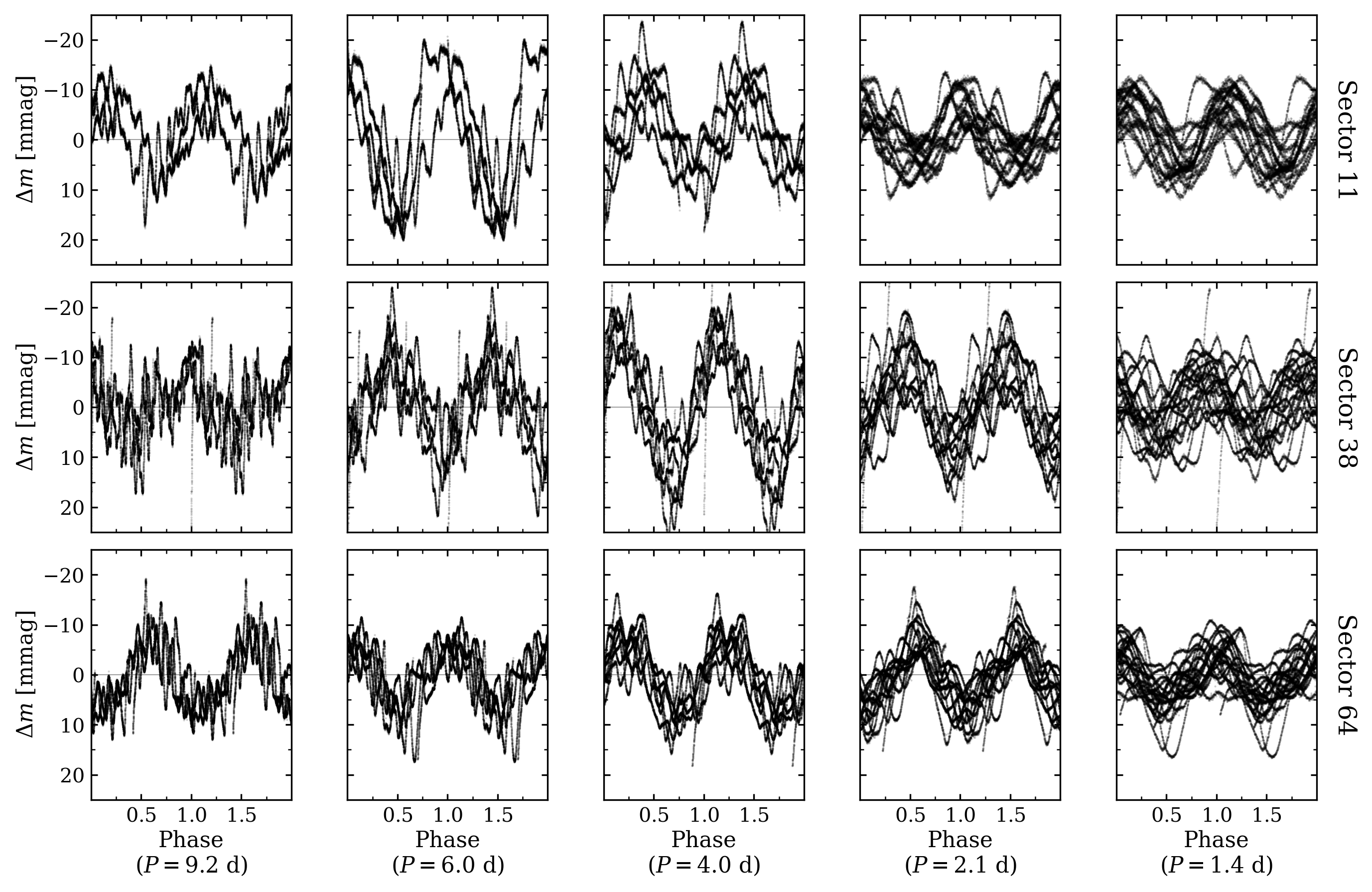}\\
    \caption{Phase-folded diagram for the independent frequencies found in \object{HD 115842} after pre-whitening. Rows: Sectors $11$, $38$, and $64$. Columns: Frequencies $f_1$--$f_5$, as listed in Table~\ref{table:all}. The average period listed in Table~\ref{tab:periods} is given below each column as a representative value.}
\label{fig:HD115842_phase}
\end{figure*}
\vskip 0.1cm

\object{HD~80077} (B2\,Ia+e) has been identified as an extreme supergiant, potentially a member of the open cluster Pismis~$11$, located at a distance of $3.2$ kpc. With an absolute bolometric magnitude of $-10.4 \pm 0.5$, this star ranks among the brightest B-type supergiants in our Galaxy \citep{Knoechel1982}. If this classification is accurate, the star should exhibit substantial instability, as it lies above the Humphreys-Davidson limit \citep{Humphreys1979}. However, \citet{Carpay1992} reported that this star appears relatively stable, and the contrast between the stability of this extreme hypergiant and its neighboring luminous blue variables (LBVs) on the HR diagram warrants further investigation. Using HIPPARCOS and V-band photometric data, \citet{Leeuwen1998} identified significant periodicities of $66.5$ days and $55.5$ days, along with additional, less prominent peaks at  $76.0$ days and $41.4$ days. \citet{Knoechel1982} observed a period of $21.2$ days through polarimetric analysis. Although no binary system has been confirmed, these authors suggested the presence of a compact companion that was originally more massive than the studied star. Additionally, \citet{ref-lefevre2009} reported light variations of $0.151$ mag over a period of $3.11$ days.

Our analysis of three TESS sector groups ($8+9$, $35+36$, and $62+63$) yields four independent frequencies. Similar to the behavior observed in the previous stars, the frequency spectrum is strongly affected by transient signals and sector-dependent amplitude variations (Fig~\ref{fig:freqHD80077}). The dominant independent frequency, $f_{1} = 0.116$\,d$^{-1}$~($P\approx8.6$\,days), displays a noticeable frequency drift between sectors. The most stable signal corresponds to $f_{2}=0.185$~d$^{-1}$~($P\approx5.4$\,days), with the highest $\mathrm{S/N}$ value in sectors~$8+9$. Two additional frequencies were identified at $f_{3}=0.248$~d$^{-1}$~($P\approx4.0$\,days) and $f_{4}=0.287$~d$^{-1}$~($P\approx3.5$\,days), the latter also showing temporal variability. Although none of the TESS frequencies match directly with the long periods reported for this object, the frequency $f_{4}$ is close to the period reported by \citet{ref-lefevre2009}. 

\vskip 0.1cm
\object{HD~99953} (B1/2\,Iab/b) is part of the Centaurus constellation.  \citet{ref-haucke-paper} reported that the star exhibits a variable P~Cygni H$\alpha$ profile, indicating that both the mass-loss rate ($\dot{M}$) and terminal wind velocity ($v_{\infty}$) vary by a factor of $\sim~2.8$, which suggests significant wind variability.

Our analysis of TESS sectors $10$, $37+38$, and $64$ reveals four independent signals between $0.11$ and $0.41$\,d$^{-1}$, corresponding to periods between approximately $2.5$ and $8.8$ days. The frequency spectrum is strongly affected by transient and low-amplitude variability.  The independent frequencies interact with each other and with harmonics, generating strong beating patterns, visible in the WWZ power spectra (Fig~\ref{fig:freqHD99953}). 

The dominant frequency, $f_{1} = 0.113$\,d$^{-1}$~($P\approx8.8$\,days), remains remarkably stable across all sectors. Similarly, $f_{2} = 0.250$\,d$^{-1}$~($P\approx4.0$\,days) and $f_{3} = 0.323$\,d$^{-1}$~($P\approx3.1$\,days) show only small sector-to-sector variations. In contrast, $f_{4} = 0.398$\,d$^{-1}$~($P\approx2.5$\,days) exhibits a frequency drift, decreasing from $\sim 0.408$ to $0.385$\,d$^{-1}$ in almost four years (see Table~\ref{table:all}).

\begin{figure*}[t!]
 \centering    
 \includegraphics[width=2.0\columnwidth]{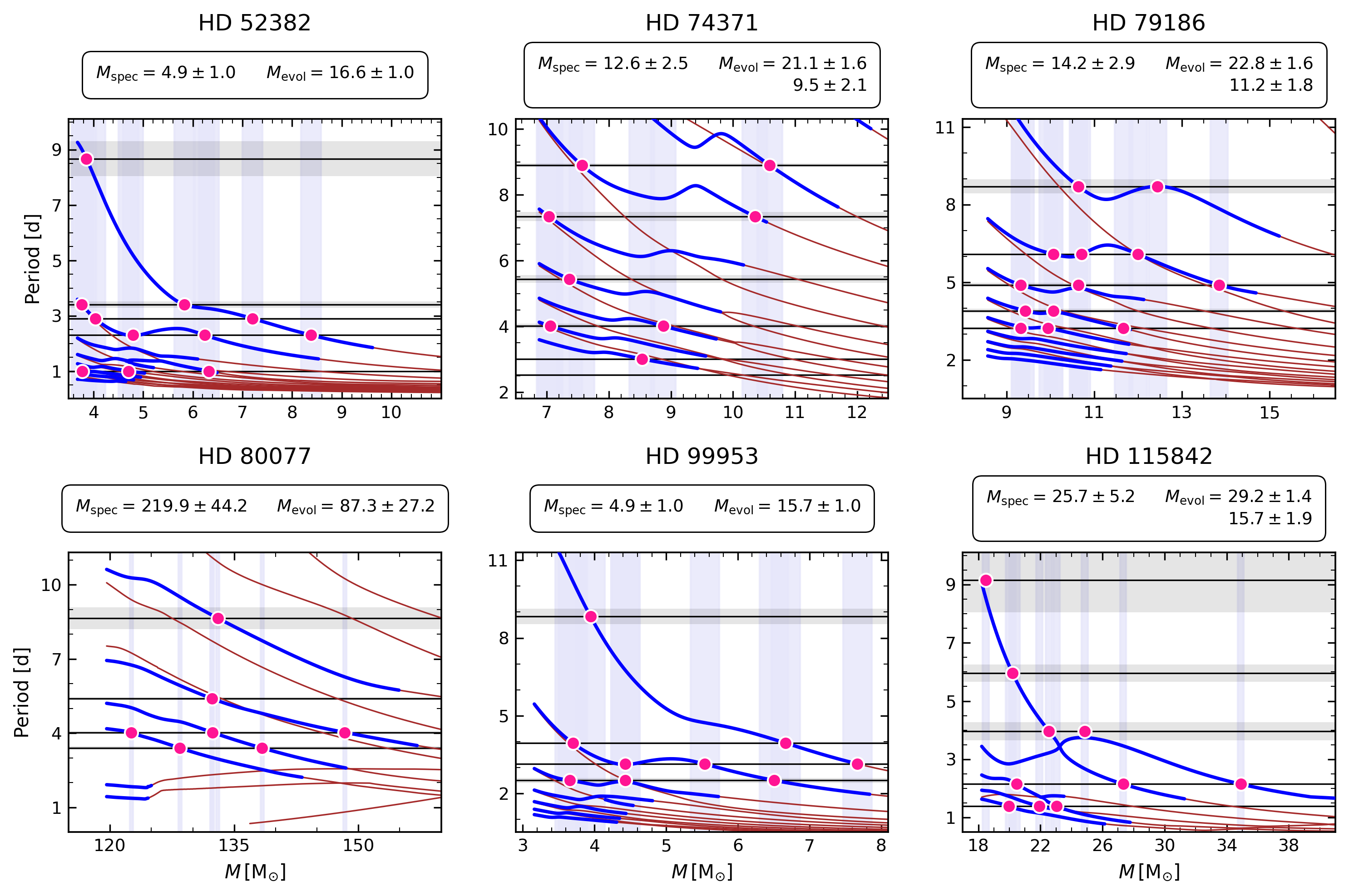} 
\caption{Period–mass modal diagram derived from the nonadiabatic pulsation models for the sample of BSGs. The thick blue lines indicate the excited pulsation modes, while the thin red lines indicate damped modes. The horizontal black lines mark the observed photometric periods, and the gray band indicates the dispersion between TESS sectors (Table~\ref{table:all}). The vertical bands indicate the mass ranges where an unstable mode intersects an observed period. We adopted a mass uncertainty of 0.5 $M_{\odot}$. The magenta circles mark the exact crossing points. The spectroscopic ($M_{spec}$) and evolutionary ($M_{evol}$) masses are included as reference values, following Table~\ref{tab:masses}, to facilitate comparison with the results shown here.}
\label{fig:modaldiagram}
\end{figure*}

The improved frequency resolution of the combined TESS sectors~$37+38$ reveals possible rotational splitting signatures. A triplet centered on $f_{2}$ is identified, with nearly symmetric separations of $\Delta f \approx 0.048$\,d$^{-1}$, corresponding to components at $f_{-} = 0.202$\,d$^{-1}$ and $f_{+} = 0.298$\,d$^{-1}$. A similar separation is observed around $f_{3}$, with additional peaks at $0.2758$ and $0.3726$\,d$^{-1}$. However, these side components are not recovered in other sectors. In addition, $f_{2}$, $f_{3}$, and $f_{4}$ also share a similar frequency separation between them of $\Delta f \approx 0.07$\,d$^{-1}$. Interestingly, all the frequencies found are very similar to those detected in \object{HD 74371}, which also exhibit the same separation of $\Delta f \approx 0.07$\,d$^{-1}$.

\object{HD~115842} (B$0.5$\,Ia/ab) is a supergiant star located in the Centaurus constellation. Photometric variations with periods of $13.38$\,days and $10.31$\,days have been reported by \citet{ref-kohen} and \citet{ref-lefevre2009}, respectively. The star also exhibits variations in the H$\alpha$ line profile \citep{ref-haucke-paper} and present a large macroturbulent velocity $V_\mathrm{mac} = 225~\mathrm{km~s^{-1}}$ \citep{Evans2004}.

Our analysis of TESS sectors~$11$, $38$, and $64$ yields five independent frequencies; all of them display amplitude or frequency variations from one sector to another. Numerous beating structures are also visible in the WWZ scalograms, as shown in Fig.~\ref{fig:freqHD115842}. The first independent frequency is $f_{1}=0.109$\,d$^{-1}$~($P\approx9.2$\,days), which varies from $0.094$ to $0.118$\,d$^{-1}$. The period detected in TESS sector~38 ($P\approx10.64$~d) is in good agreement with the $10.31$ day period reported by \citet{ref-lefevre2009}. Since the observed frequency variations remain within the Rayleigh resolution limit of a single TESS sector, as for \object{HD 52382}, $f_{1}$ is interpreted as the same oscillation mode undergoing temporal modulation. The frequencies $f_{2} = 0.168$\,d$^{-1}$~($P\approx6.0$\,days) and $f_{3}=0.252$\,d$^{-1}$~($P\approx4.0$\,days) both exhibit systematic, monotonically increasing frequency shifts between consecutive sectors. This behavior differs from the more irregular pattern expected from SLFV alone and may instead indicate long-term modulation. In contrast, the higher-frequency signals $f_{4}=0.472$\,d$^{-1}$~($P\approx2.1$\,days) and $ f_{5}=0.710$\,d$^{-1}$~($P\approx1.4$\,days) remain more stable across all sectors, making them the strongest candidates for coherent pulsation modes. To further assess the stability of the detected variability, we constructed phase-folded light curves, shown in Fig.~\ref{fig:HD115842_phase}, for each independent frequency after pre-whitening all remaining independent signals.

\begin{table*}[ht]
\caption{Different masses obtained for the sample of stars.}
\centering
\renewcommand{\arraystretch}{1.15}
\resizebox{\textwidth}{!}{%
\begin{threeparttable}
\setlength{\tabcolsep}{5pt}
\begin{tabular}{l c c  c c  c c  c c  c c  c c}
\toprule
\toprule
&
  \multicolumn{2}{c}{HD~52382} &
  \multicolumn{2}{c}{HD~74371} &
  \multicolumn{2}{c}{HD~79186} &
  \multicolumn{2}{c}{HD~80077} &
  \multicolumn{2}{c}{HD~99953} &
  \multicolumn{2}{c}{HD~115842} \\
\midrule
&
  $M\pm\sigma$ & $N_P/T$ &
  $M\pm\sigma$ & $N_P/T$ &
  $M\pm\sigma$ & $N_P/T$ &
  $M\pm\sigma$ & $N_P/T$ &
  $M\pm\sigma$ & $N_P/T$ &
  $M\pm\sigma$ & $N_P/T$ \\
& [$M_\odot$] & & [$M_\odot$] & & [$M_\odot$] & & [$M_\odot$] & & [$M_\odot$] & & [$M_\odot$] & \\
\midrule
\multirow{5}{*}{\shortstack{$M_{ast}$}}
  & $3.90\pm0.38$   & $3/5$ & $7.30\pm0.52$   & $4/6$ & $9.37\pm0.30$   & $3/5$ & $122.58\pm0.25$ & $1/4$ & $3.80\pm0.37$ & $3/4$ & $18.49\pm0.25$ & $1/5$  \\
  & $4.75\pm0.25$   & $2/5$ & $8.70\pm0.42$   & $2/6$ & $10.06\pm0.27$  & $3/$5 & $128.45\pm0.25$ & $1/4$  &  $4.43\pm0.25$ & $2/4$ &  $20.08\pm0.30$ & $3/5$ \\ 
  & $6.08\pm0.29$   & $3/5$ & $10.35\pm0.37$  & $2/6$ & $10.67\pm0.29$  & $3/5$ &  $132.66\pm0.35$ & $3/4$ & $5.54\pm0.25$ & $1/4$ & $22.49\pm0.25$ & $2/5$ \\ 
  & $7.20\pm0.25$   & $1/5$ & & & $12.09\pm0.60$ & $3/5$ & $138.37\pm0.25$ & $1/4$ & $6.59\pm0.25$  & $2/4$ & $24.86\pm0.25$ & $1/5$ \\
  & $8.38\pm0.25$   & $1/5$ & &  & $13.85\pm0.25$ & $1/5$ & $148.37\pm0.25$ & $1/4$ &$7.67\pm0.25$  & $1/4$ &  $27.33\pm0.25$ & $1/5$ \\ 
  & & & & & & & & & & & $34.90\pm0.25$ & $1/5$ \\
\midrule
\shortstack{$M_{spec}$}
  & \multicolumn{2}{c}{$4.86\pm0.98$}
  & \multicolumn{2}{c}{$12.56\pm2.54$}
  & \multicolumn{2}{c}{$14.22\pm2.86$}
  & \multicolumn{2}{c}{$219.92\pm44.23$}
  & \multicolumn{2}{c}{$4.88\pm0.98$}
  & \multicolumn{2}{c}{$25.72\pm5.17$} \\
\midrule
\multirow{2}{*}{\shortstack{$M_{evol}$}}
  & \multicolumn{2}{c}{$16.62\pm 0.96$}
  & \multicolumn{2}{c}{$21.12\pm 1.64$}
  & \multicolumn{2}{c}{$22.77\pm 1.63$}
  & \multicolumn{2}{c}{$87.26\pm 27.22$}
  & \multicolumn{2}{c}{$15.66\pm 0.97$}
  & \multicolumn{2}{c}{$29.17\pm 1.44 $} \\
  & &
  & \multicolumn{2}{c}{$9.52\pm 2.08$}
  & \multicolumn{2}{c}{$11.23\pm 1.78$}
  & &
  & &
  & \multicolumn{2}{c}{$15.73\pm 1.87$}
  \rule{0pt}{1em}\\
\bottomrule
\end{tabular}
\textbf{Note}: Asteroseismic mass ranges ($M_{ast}$) are listed together with the fraction of matched periods over the total stable periods observed ($N_P/T$). The spectroscopic mass ($M_{spec}$) is given for comparison. The evolutionary mass ($M_{evol}$) corresponds to the position of the star on the HR diagram. Two values are provided when the star lies in a position where the blue loop can occur. Dispersions for $M_{evol}$ were propagated from luminosity errors. For $M_{ast}$, the uncertainty corresponds to half the adopted mass band ($\pm0.25$~M$_\odot$) used to identify period matches, extended by the spread of crossings within each range.
\end{threeparttable}
}
\label{tab:masses}
\end{table*}

\subsection{Linear nonadiabatic analysis}\label{sec:modelsequence}

The resulting periods for all the objects and their associated nonadiabatic model sequences are presented in Fig.~\ref{fig:modaldiagram} as a period-mass ($P-M$) diagram, in terms of the lowest-order pulsation periods. At high masses, the eigenfrequency spectrum is well ordered and resembles that of a classical acoustic resonator (thin red lines). As the stellar mass decreases, or equivalently, as the $L/M$ increases, mode crossings become more frequent and give rise to the appearance of strange modes (thick blue lines). These unstable modes are characteristic of highly nonadiabatic stellar envelopes and are associated with increasingly violent instabilities, particularly when $L/M > 10^4 $ \citep{Glatzel1994}. 

As shown in Fig.~\ref{fig:modaldiagram}, all stars in our sample exhibit unstable strange modes with periods consistent with those observed (magenta circles). This diagram is used to constrain the current stellar mass (i.e., the asteroseismic mass). However, the solutions are not unique, and the possible asteroseismic mass intervals can be estimated when the observed periods (horizontal lines) intersect the loci of unstable modes within a narrow mass interval. Conversely, for a stellar mass derived from independent methods, the predicted periods of unstable modes can be compared with the observed periods, allowing the identification and classification of the detected pulsations. 

In this work, we used the observed periods to estimate the stellar mass. We adopted uncertainties of about $0.5$~M$_\sun$, which represent the spread of acceptable solutions (vertical bands) defined by the uncertainty in the periods (horizontal bands). In each case, the intersections cluster around a common mass value, producing an approximately vertical alignment in the diagram. This approach assumes that the periods predicted by the linear pulsation analysis are equal to the observed periods. Although this approximation is generally valid for moderately unstable modes, nonlinear effects may become important for modes with dynamical growth rates, particularly at the low-mass end of the model sequences. In such cases, the instabilities tend to inflate the envelopes when evolving into the nonlinear regime, \citep[see, e.g.,][]{Yadavetal2021} and, therefore, the final pulsation periods are longer than their linear counterparts. Consequently, the mass inferred from the observed periods in a linear analysis should be considered as a first approximation to the asteroseismic mass of the star. 

The resulting mass estimates corresponding to the different solutions are listed in Table~\ref{tab:masses}. For each mass interval (vertical bands), we adopted the mean mass. When different periods yielded overlapping mass intervals, they were assumed to correspond to the same solution and were grouped into a single mass range. The subsequent column indicates the number of observed periods, relative to the total number of detected periods used in the analysis, successfully matched by the unstable modes.

For none of the stars does a single mass range fully explain all detected periods, indicating either that the asteroseismic solution is not unique or that multiple physical processes may contribute to the observed variability. For \object{HD~52382} and \object{HD~99953}, the asteroseismic mass ranges are similar, yielding several discrete mass solutions between $3.8$ to $8.4$ M$_\odot$. The masses that cluster around the largest number of detected periods correspond to $3.9\pm0.4$~M$_\odot$ and $6.1\pm0.3$~M$_\odot$ for \object{HD 52382}, and $3.8\pm0.4$~M$_\odot$ for \object{HD~99953}.

\object{HD~74371} and \object{HD~79186} display higher asteroseismic masses, between $7.3$ to $10.4$~M$_\odot$ and $9.4$ to $13.9$~M$_\odot$, respectively. For \object{HD~74371}, five of the six observed periods are associated with different sequences of unstable modes, whereas the remaining period does not intersect any unstable mode. A mass of $7.3\pm 0.5$~M$_\odot$ can account for the majority of observed periods. For \object{HD~79186}, four discrete mass values are equally probable between $9.4$ and $12.1$~M$_\odot$, leading to a mean value of $10.7\pm 1.4$~M$_\odot$. Masses between $122$~M$_\odot$ to $148$~M$_\odot$ are predicted for \object{HD~80077}, being the most probable value of $132.7\pm0.4$~M$_\odot$, since it reproduces three of the four observed periods. 

For the last object in the sample, \object{HD 115842}, the radial nonadiabatic models predict several possible masses between $18$ and $35~$M$_{\odot}$, depending on the observed periods considered. Among all the solutions, the mass of $20.1 \pm 0.3$~M$_{\odot}$ reproduces three of the five observed periods.

\section{Discussion}\label{sec:5}

\subsection{Comparison of mass determinations and evolutionary implications}

The mass discrepancy problem in massive stars was first identified by \citet{Herrero1992}, who performed a non-local thermodynamic equilibrium (NLTE) analysis of $25$ Galactic O stars and found that spectroscopic masses were systematically lower than evolutionary masses. These masses are derived by comparing the position of the star on the HR diagram with theoretical evolutionary tracks. The discrepancies in both mass and helium abundances are significant, pointing to a failure in either the atmosphere or the stellar evolution models, suggesting that a higher mass loss rate at the beginning of the evolution should be adopted in the calculations to reduce the discrepancy. In subsequent decades, the inclusion of line-blanketed atmospheres, spherical stellar winds, and rotating evolutionary models substantially reduced the magnitude of the problem, though without fully resolving it \citep{Crowther2006, Searle2008}, particularly for stars below $30\, M_\odot$ \citep{Markova2018}. Moreover, \cite{Sanchez2023} suggest that the mass discrepancy problem may be part of a broader tension between spectroscopic diagnosis and evolutionary predictions. They show that even when stellar masses and luminosities can be quantitatively matched by evolutionary models, the observed CNO surface abundances may remain inconsistent with evolutionary predictions and highlight the need for large, homogeneously analyzed samples combining multi-epoch spectroscopy and long-baseline photometry. Additional recent studies have also linked the origin of this discrepancy to convective core boundary mixing, suggesting that evolutionary models require more massive helium cores than previously assumed to reproduce spectroscopic masses \citep{Tkachenko2020, Pedersen2021}.

To evaluate a possible mass discrepancy in this sample of BSGs and establish a comparison with the asteroseismic mass estimates, we derived the spectroscopic masses from the stellar radii and surface gravities reported in Table~\ref{table:observations}:

\begin{equation}
    M_\mathrm{spec}\, = \frac{g_{corr}\,R^2}{G}\,.
\end{equation}
The resulting spectroscopic masses are listed in Table~\ref{tab:masses}. 

\begin{figure}[t!]
 \centering
    \includegraphics[width=0.95\columnwidth]{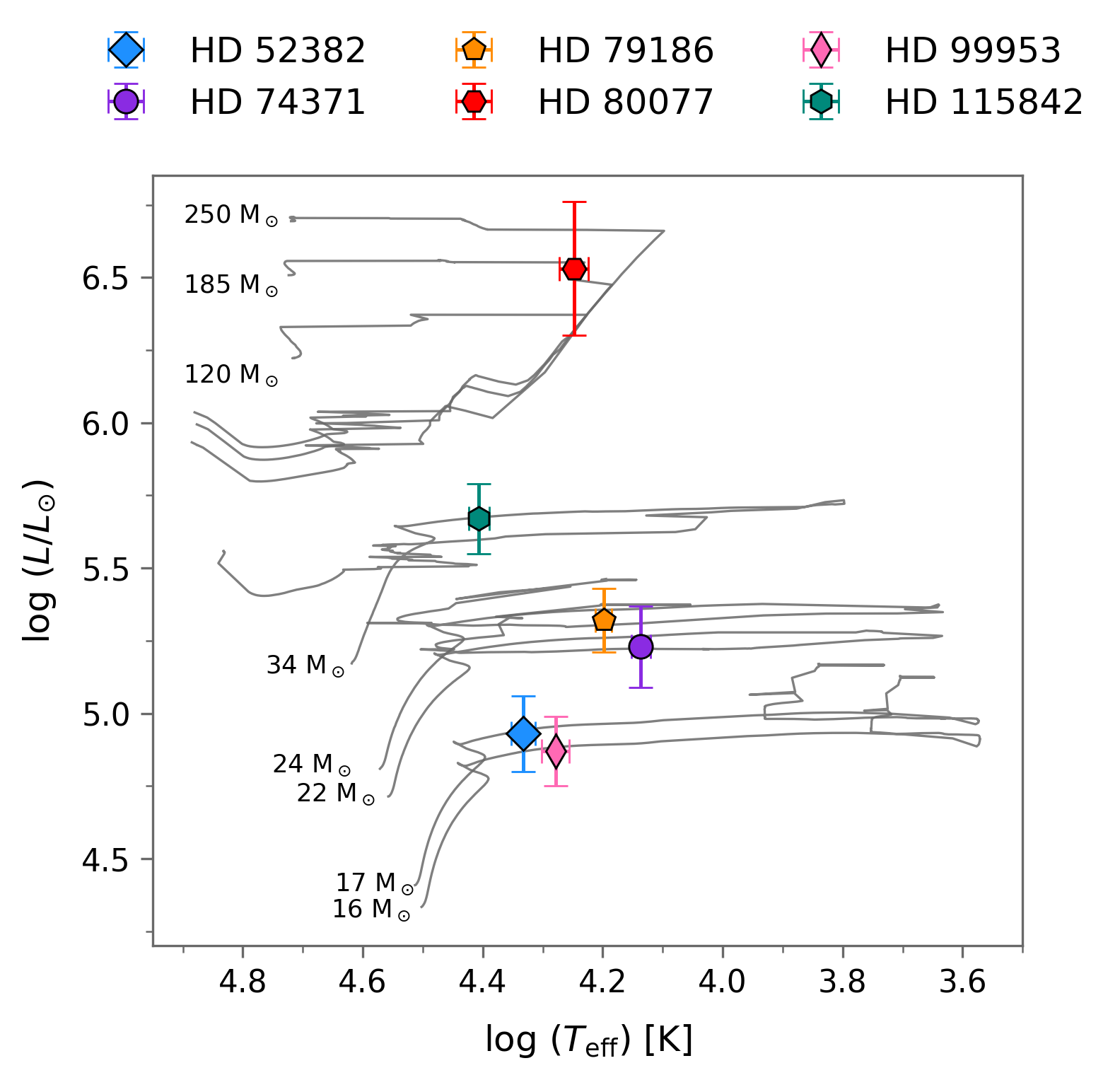}    
 \caption{Position of the sample of BSGs on the HR diagram. The evolutionary tracks, considering rotation $\Omega / \Omega_{crit} = 0.4$ and solar metallicity Z=0.014, were taken from \cite{ref-ekstr2012}.} 
 \label{fig:HRdiagram}
\end{figure}

To estimate the evolutionary masses, we interpolated the position of each star on the HR diagram into the nearest evolutionary tracks, using the rotating stellar evolution models computed with the Geneva stellar evolution code \citep{ref-ekstr2012} (see Fig~\ref{fig:HRdiagram}). These models cover a range of initial masses from $16$ to $250$\,M$_\sun$ at solar metallicity ($Z = 0.014$), with a ratio between the equatorial surface rotational velocity and critical rotational velocity of $\Omega= 0.4$. The obtained values are listed in Table~\ref{tab:masses}. 

For \object{HD~52382} and \object{HD~99953}, the evolutionary tracks on the HR diagram suggest initial masses of $17$ and $16$\,M$_{\odot}$, respectively (see Fig.~\ref{fig:HRdiagram}), leading to a current stellar mass of the same order ($16.6$ and $15.7$~M$_{\odot}$). In contrast, the derived asteroseismic ($4$~M$_\sun$ and $6$~M$_\sun$) and spectroscopic masses ($5$ M$_\sun$) are in good agreement with each other.

\citet{ref-haucke-paper} mention that these objects show signs of interaction with the interstellar medium (see their Figs.~2 and~4). This observation, along with several modes that cross each other in the modal diagram as well as low spectroscopic mass estimates, may indicate that both stars are returning from the RSG phase, where they experienced shedding of material into their surroundings. As the adopted models cease to show a blueward evolution at that low-mass regime, possible explanations to support this scenario include enhanced mass-loss rates during the RSG stage  \citep{Georgy2012, Saio2013} and erroneous distance estimates.

For \object{HD 74371} and \object{HD 79186}, their positions on the HR diagram correspond to evolutionary tracks with initial masses of approximately $22$\,M$_{\odot}$ and $24$\,M$_{\odot}$, respectively (see Fig.~\ref{fig:HRdiagram}). At these locations, the stars may either be evolving toward the RSG phase or returning to the BSG phase.

When comparing the most probable asteroseismic mass of $7.3\pm0.5$~M$_\odot$ derived for \object{HD 74371} with the spectroscopic mass ($12.6\pm2.5$~M$_{\odot}$) and the current predictions of evolutionary tracks (listed in Table~\ref{tab:masses}), we find agreement with a post-RSG scenario, although some discrepancies remain. For \object{HD~79186}, the average asteroseismological mass of $10.7\pm 1.4$~M$_\odot$ closely aligns with the evolutionary mass of $11.23$~M$_{\odot}$ expected in a post-RSG phase. This is also consistent with the spectroscopic estimate of $14.2\pm2.8$~M$_\odot$, within the uncertainties. Although \object{HD~79186} was not previously classified as an $\alpha$~Cygni variable, such a classification is plausible given the observed variability and the mass ranges inferred from this analysis.

For \object{HD~115842}, evolutionary models indicate that the track with an initial mass of approximately $34$~M$_{\odot}$ best matches the position of the star on the HR diagram (see Fig.~\ref{fig:HRdiagram}). The current predicted mass is either $29.2$~M$_{\odot}$ or $15.7$~M$_{\odot}$, depending on whether the star is in a pre- or post-RSG stage, respectively. The spectroscopic mass of $25.7 \pm 5.2$~M$_{\odot}$ is consistent, within the uncertainties, with the pre-RSG evolutionary mass. The most probable asteroseismic estimate of $20.1$~M$_{\odot}$ is closer to the post-RSG evolutionary mass and is also consistent, within the uncertainties, with the spectroscopic mass. Therefore, both the pre- and post-RSG scenarios remain compatible within the available constraints.

In addition, according to \citet{ref-haucke-paper}, combined \textit{WISE} W3 and W4 images reveal a prominent bow-shock structure (see their Fig.~6), indicating a strong interaction between the stellar wind and the interstellar medium. Furthermore, \cite{Crowther2006, Searle2008} reported that this star exhibits a nearly normal CNO abundance pattern for its spectral type, with a slight nitrogen overabundance. Taken together, these observational constraints suggest that the star has undergone significant mass loss through stellar winds after leaving the MS. \\

\begin{figure*}[t!]
    \centering
    \includegraphics[width=0.95\linewidth]{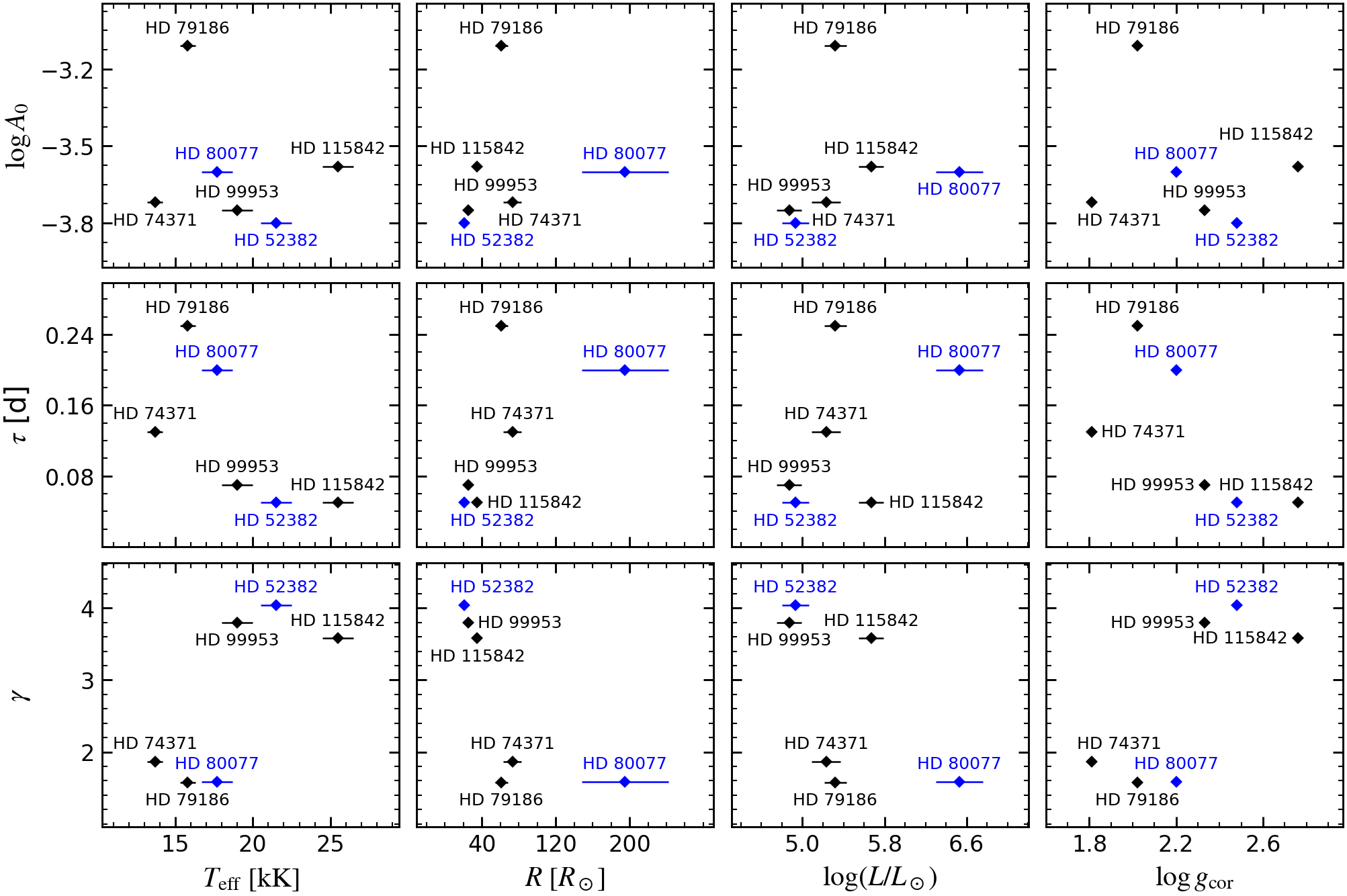}
    \caption{SLFV parameters $\log A_0$, $\tau$, and $\gamma$ as functions of the stellar parameters $T_{\rm eff}$, $R$, $\log(L/L_\odot)$, and $\log g_{\rm cor}$ for the BSG sample. The blue-filled diamonds correspond to stars analyzed in this work, while the black-filled diamonds indicate values taken from \citet{Kourniotis2025}. The horizontal error bars reflect the uncertainties in the stellar parameters.}
    \label{fig:slfv_vs_stellarp}
\end{figure*}

\subsection{The peculiar case of the hypergiant HD 80077}

This star exhibits the largest mass discrepancies of the entire sample. Its position on the HR diagram identifies it as a very massive post-MS star near the turning point of its evolutionary track before evolving back toward the hot side of the HR diagram. At that point, the star reaches its minimum effective temperature, placing it on the cool side of the bistability jump associated with the Fe recombination, where a denser and slower wind is expected.
The evolutionary track most consistent with its position corresponds to an initial mass of $185$~M$_{\odot}$ (see Fig.~\ref{fig:HRdiagram}), leading to a current predicted mass of approximately $87$~M$_{\odot}$. Instead, the asteroseismological model yields a stellar mass of about $133$~M$_{\odot}$ from the simultaneous fit of three observed periods. The spectroscopic value is even larger, reaching $220$~M$_{\odot}$, which is almost $2.5$ times the evolutionary mass. While part of the discrepancy may be attributed to the large uncertainty in the luminosity arising from the poorly constrained distance, the adopted mass-loss prescription is another significant source of uncertainty, particularly for such highly evolved massive stars. For example,  \citet{Vink2000, Vink2001} predict an increase in the mass-loss rate across the bistability jump, while \citet{ref-haucke-paper} observe a decrease and estimated a mass-loss rate of $5.4\times10^{-6}$~M$_\odot\mathrm{yr}^{-1}$ for this object.

As the disagreement among the independent mass estimates remains significant, such large discrepancies may indicate that the underlying physics of this object is more complex than that of a single evolving star.  Despite the absence of a previously reported binary companion and the lack of evidence for binarity in the TESS data analyzed in this work, this object warrants a more detailed analysis. Particularly, determinations of helium abundances, and $N/C$ and $N/O$ surface ratios would provide strong constraints on the degree of internal mixing and mass-loss history.

\subsection{SLFV parameters and their relation to stellar properties}

Recent studies have revealed correlations between the SLFV parameters and the stellar properties of massive stars. For early-type stars spanning the upper MS, a negative correlation between the red-noise amplitude ($A_0$) and $T_{\mathrm{eff}}$ has been reported \citep{Bowman2020, Shen2024}. An inverse association between these two parameters was found by \citet{Kourniotis2025} in a sample of $41$ Galactic post-MS BSGs. Taken together, these results suggest that the SLFV may be amplified at or shortly after the end of the MS phase. Moreover, for BSGs, the characteristic timescale $\tau$ is found to be negatively correlated with $\log T_{\rm eff}$, while the slope $\gamma$ correlates most strongly with $\log(L/L_\odot)$, after removing the $\alpha$~Cygni variables \citep{Kourniotis2025}. In addition, \citet{Spejcher2025} report no significant correlations between $A_0$ and the stellar properties for a sample of LBVs and high-luminosity BSGs, suggesting that these stars may be an extension of the $\alpha$~Cygni variables. 

To test whether these trends persist in our sample, we compared the SLFV parameters of our six B supergiants spanning both pre- and post-RSG evolutionary stages, all of which exhibit strange modes, as a function of their stellar properties. In Fig.~\ref{fig:slfv_vs_stellarp}, we show the SLFV parameters $\log A_0$, $\tau$, and $\gamma$ as functions of the stellar parameters $T_{\rm eff}$, $R$, $\log(L)$, and $\log g_{\rm cor}$. The two stars analyzed in this work (\object{HD 52382} and \object{HD 80077}) are highlighted in blue, whereas the remaining four stars, whose SLFV properties were adopted from \citet{Kourniotis2025}, are shown in black diamonds.

No relation is seen between $\log A_0$ and any of the stellar parameters, with \object{HD~79186} appearing as a persistent outlier with respect to $\log A_0$. For $\tau$, a tentative positive trend with $R$ and a negative trend with $T_{\rm eff}$ are observed when \object{HD~79186} is excluded, suggesting that more extended and cooler supergiants tend to exhibit longer variability timescales. Regarding $\gamma$, the sample reveals a clear bimodal structure: the hotter, more compact supergiants (\object{HD~52382}, \object{HD~99953}, and \object{HD~115842}) exhibit steep power-law slopes, while the cooler, more extended objects (\object{HD~74371}, \object{HD~79186}, and \object{HD~80077}) show flatter slopes.

For none of the stars do the SLFV parameters show a clear segmentation between the post and pre-RSG candidates, suggesting that the SLFV properties alone are not sufficient to distinguish evolutionary stages within this sample. As seen in  \cite{Kourniotis2025}, no trends with the SLFV properties are observed in the $\alpha$~Cygni class. These results support the idea that $\alpha$~Cygni variables do not constitute a homogeneous class, neither in evolutionary state nor in the physical origin of their variability \citep{Saio2013, Kourniotis2025, Spejcher2025}. We note that the derived SLFV parameters are sensitive to the methodology employed \citep{Krticka2021, Ma2024, Pedersen2025}. Model-independent tests, as proposed by \citet{Pedersen2025}, may provide complementary constraints.

\section{Conclusions}\label{sec:6}

We analyzed TESS light curves of six BSG stars with a large $L/M$ to identify periods potentially associated with strange modes. Using hydrostatic stellar models, we tested their envelope structures for linear nonadiabatic stability under radial perturbations and compared the resulting theoretical pulsation periods with the observational data. This approach allowed us to estimate the most likely asteroseismic masses of either $3.9$~M$_{\odot}$ or $6.1$~M$_{\odot}$ for \object{HD~52382}, $7.3$~M$_{\odot}$ for \object{HD~74371}, $10.7$~M$_{\odot}$ for \object{HD~79186}, $133$~M$_{\odot}$ for \object{HD~80077}, $3.8$~M$_{\odot}$ for \object{HD~99953}, and $20.1$~M$_{\odot}$ for \object{HD~115842}. These values were then compared with current evolutionary masses derived from the Geneva stellar models \citep{ref-ekstr2012} and with spectroscopic masses obtained according to the stellar radii and $\log g_{corr}$ estimates from \cite{ref-haucke-paper}. From this simultaneous analysis, we find that \object{HD~74371} and \object{HD~79186} are evolving blueward following the RSG phase, while the evolutionary stage of \object{HD~115842} cannot be uniquely determined, with both pre- and post-RSG scenarios compatible with the available mass estimates. We also find that \object{HD~52382} and \object{HD~99953} are likely in the post-RSG phase; however, their low masses are difficult to reconcile with the available evolutionary models. On the other hand, \object{HD~80077} lies close to the turning point of its evolutionary track, exhibiting the largest mass discrepancy in our sample. This indicates that the underlying physics of this object is more complex than that of a single evolving star and warrants a more detailed analysis.

Overall, the SLFV parameters do not provide a clear diagnostic of the evolutionary stage of the stars of our sample, as no systematic trends were found. Nevertheless, the observed bimodal distribution in the characteristic timescales and power-law slopes suggests that the physical mechanisms driving the low-frequency variability may differ among $\alpha$~Cygni variables. Expanding this analysis to a larger sample will be essential to establish whether these trends are intrinsic to the class or reflect the limited size of the current sample.

In summary, we highlighted the potential of asteroseismology as a complementary tool for constraining the evolutionary status of BSGs. However, the mass discrepancies found for some objects indicate that further observational and theoretical efforts are still required. Moreover, since not all of the photometric periods identified in the TESS light curves find a counterpart in our theoretical radial pulsation models, an analysis of non-radial pulsations could help elucidate their physical origin and further characterize the rich variability observed in $\alpha$~Cyg stars. Such studies constitute important next steps toward a more complete understanding of the variability and evolution of BSGs.

\begin{acknowledgements}
 
   This research work has been co-funded by the European Union, Project 101183150 - OCEANS and by the Program for Research and Innovation Horizon 2020 (2014-2020) under the Marie Sk\l{}odowska-Curie Grant Agreement No. 823734. AA and LC thank the financial support from the Universidad Nacional de La Plata (Programa de Incentivos 11/G160), Argentina. MK and JSA acknowledges financial support from the Czech Sciencie Foundation (proyect GA\v{C}R 25-17532S). The Astronomical Institute Ondřejov is supported by RVO:67985815. AC acknowledge support from Centro de Estudios Atmosféricos y Astroestadística (CEAAS), Universidad de Valparaíso, Chile.
    
\end{acknowledgements}

\bibliographystyle{aa} 
\bibliography{Biblio}

\appendix
\section{Frequency analysis}\label{appexA}

\subsection{Lomb-Scargle periodogram}

We performed a Lomb-Scargle periodogram analysis  \citep{ref-lomb, Scargle1982} using the \texttt{timeseries.LombScargle} package from the \textsc{Python} module \textsc{Astropy} \citep{astropy:2013, astropy:2018, astropy:2022}. The resulting power spectral density (PSD) was converted into amplitude ($A$) using the relation \(A(f) = \sqrt{PSD(f)/(N/4)}\), where $f$ is the frequency and \( N \) is the number of data points in the light curve \citep{Deeming1975, Scargle1982, Naze2021}. Significant frequencies were extracted through an iterative pre-whitening process. At each iteration, we selected the frequency with the highest amplitude, fitted the light curve with the corresponding sinusoidal model, and subtracted it. The residuals were then subjected to a new iteration to identify the next frequency. This process was repeated until no frequency exceeded the significance criterion. \citet{Baran2021} recommended a detection threshold of $\mathrm{S/N} \ge 5$ for TESS data to minimize the risk of spurious peaks. However, as these objects are strongly affected by SLFV, we adopted the traditional $\mathrm{S/N} \ge 4$ here. The signal-to-noise ratio was computed as the ratio between the amplitude of the extracted frequency and the mean amplitude of the residuals within a symmetric frequency window centered on the peak, excluding a narrow region around the peak itself. \citet{Burssens2020} proposed using a $1~\mathrm{d}^{-1}$ spectral window when significant SLFV is present in TESS data, to prevent the enhanced low-frequency noise from biasing the S/N estimate. When coherent pulsations coexist with SLFV, they suggested a broader $5~\mathrm{d}^{-1}$ window to obtain a more representative noise level, although there is no strict consensus on this choice. We therefore adopted an intermediate value of $2.5~\mathrm{d}^{-1}$ for each target.

Frequency uncertainties ($s_{f}$) were estimated following \citet{Kallinger2008} as
\begin{equation}
\begin{aligned}
    s_{f} &= \frac{1}{T\,\sqrt{sig}},\  \hskip 0.3cm \text{and}\  \hskip 0.3cm
    sig = \frac{\pi \log_{10} e}{4} \left(\frac{S}{N}\right)^2.
\end{aligned}
\end{equation}
\noindent 
where $T$ is the total time span of the observations and \textit{sig} is the spectral significance approximated from the signal to noise ratio \citep{Reegen2007}. This formulation provides an upper bound on the frequency uncertainty, in contrast to the commonly used expression of \citet{Montgomery1999}, which is derived under the assumption of white, uncorrelated noise and therefore represents a formal lower limit. In practice, \citet{Kallinger2008} showed through numerical simulations that approximately 30\% of cases exceed the \citet{Montgomery1999} estimate, and that the true uncertainty is further inflated in the presence of correlated noise. Given that the amplitude spectra of BSGs are characterized by SLFV, we adopted the \citet{Kallinger2008} expression as a more conservative and physically motivated choice. Period uncertainties were propagated as $s_{P} = s_{f}/f^{2}$. The detected frequencies, derived periods, S/N values, and uncertainties for each object are listed in Table~\ref{table:all}.

\subsection{Weighted wavelet Z-transform (WWZ)}

For the wavelet analysis, we employed the \textsc{libwwz} package available in \textsc{Python}. The wavelet transform provides information on the time evolution of the parameters (frequency, period, amplitude, and power) used to describe periodic and quasiperiodic signals. However, time-averaged values extracted from a wavelet analysis can be misleading due to the high sensitivity of the method to phenomena such as beating between close frequencies or multi-signal interference \citep{Torrence1998, ref-bravo2014, Alberici2023, Vitali2026}. For this reason, we relied exclusively on the WWZ scalogram to discuss the temporal behavior of the detected frequencies.

\subsection{Selection criteria}

Following the methodology employed by \cite{Kourniotis2025}, the extracted frequencies were categorized into four types using the Rayleigh resolution $1/\Delta T$, where $\Delta T$ is the time span of the data:

\begin{enumerate}
    \item Independent frequencies, $f_i$. These are significant peaks physically interpreted as real signals, which could be either oscillation modes (g- or p-modes), surface spots, or orbital periods. 

    \item Harmonic frequencies, identified as integer multiples of the independent ones. Harmonics in light curves arise from their non-sinusoidal nature and reflect the nonlinear response of the stellar medium to the propagation of pulsation waves. Consequently, they are not classified as intrinsic pulsation modes and are presumed to replicate the characteristics of their independent parent modes. Strong harmonics are therefore a consequence of nonlinear pulsations, as expected in strange modes, but they also appear in high-amplitude non-radial pulsators. Also, harmonics can appear as a result of rotational modulation, as seen for massive stars when r-modes are present \citep{Li2019}.

    \item Combination frequencies, defined as algebraic sums of two previously extracted frequencies. Separately, a frequency was flagged as a linear combination of others if the difference between the observed and predicted frequencies fell within the same $1/\Delta T$ criterion.

    \item Unresolved frequencies: Two frequencies were considered unresolved if the difference between them did not exceed the $1/\Delta T$ spectral resolution \citep{Loumos1978}.   
\end{enumerate}

It is important to note that after identifying two or three independent frequencies, most of the remaining frequencies can be explained as linear combinations. To avoid discarding potentially real signals that may accidentally satisfy the Rayleigh criterion and, therefore, be misclassified as combination frequencies or unresolved components, we searched for additional evidence by constructing phase diagrams. In this work, we define independent frequencies as physically meaningful oscillations. Therefore, frequencies were classified as independent only if they were detected in all sectors or, at least, in two of three nonconsecutive sectors, and they exhibit an unambiguous structure in their phase diagrams. The independent frequencies under this criterion are listed in Table~\ref{table:all}.

\begin{figure*}
    \centering
    \includegraphics[width=0.33\linewidth]{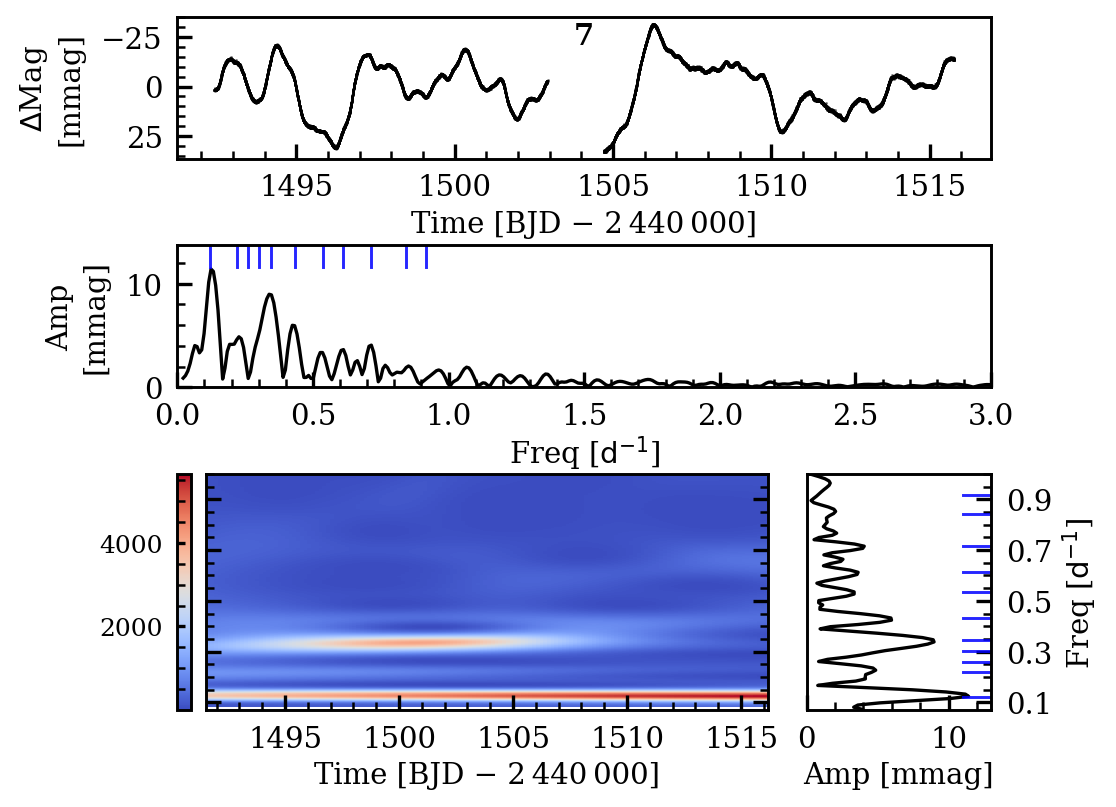} 
    \includegraphics[width=0.33\linewidth]{Graphics/HD52382_sec33.png}
    \includegraphics[width=0.33\linewidth]{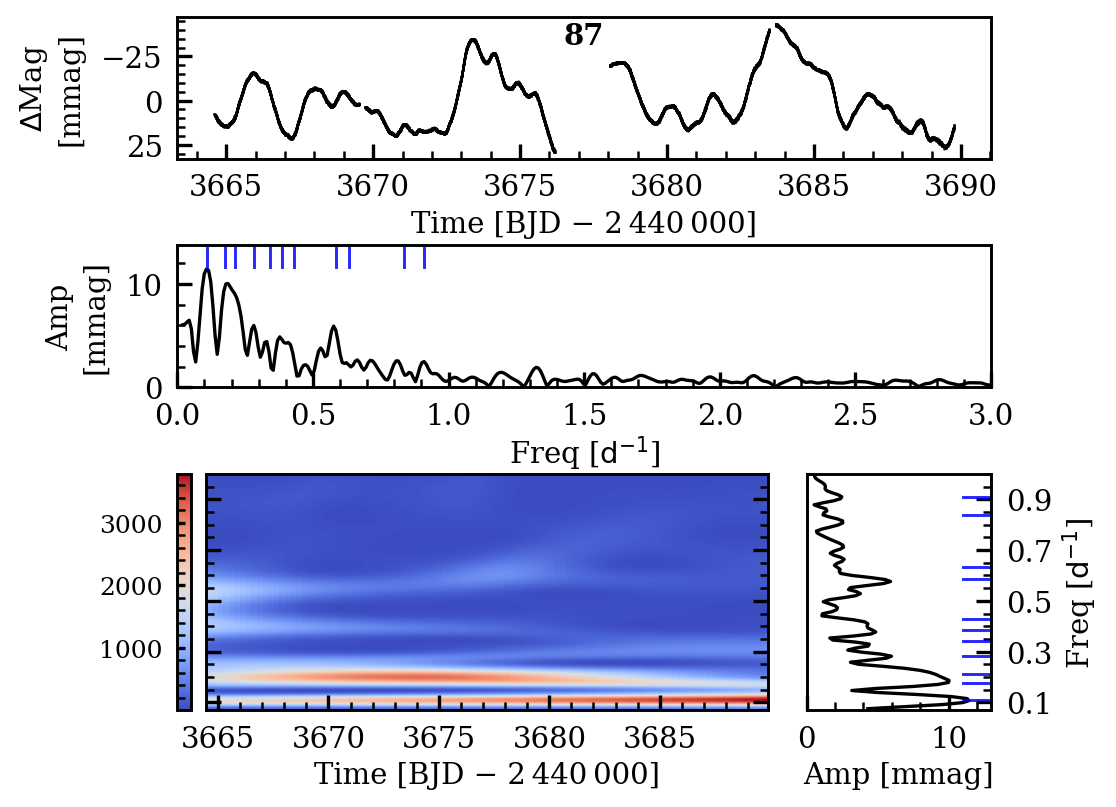}
    \caption{Frequency analysis of \object{HD~52382} from TESS sectors $7, 33$, and $87$. {\it Top panels:} TESS light curves showing the differential magnitudes as functions of time. {\it Middle panels:} Lomb-Scargle periodograms, showing the amplitude spectrum up to $3~\mathrm{d^{-1}}$. The vertical blue tick marks at the top of the plots indicate the frequencies extracted via iterative pre-whitening. {\it Bottom panels:} {\it Left:} Weighted WWZ time-frequency maps, with the color scales indicating the local power. {\it Right:} Zoom-in view on the Lomb-Scargle periodograms, shown on the same frequency scales as the scalograms.}
    \label{fig:freqHD52382}
\end{figure*}

\begin{figure*}
    \centering
    \includegraphics[width=0.33\linewidth]{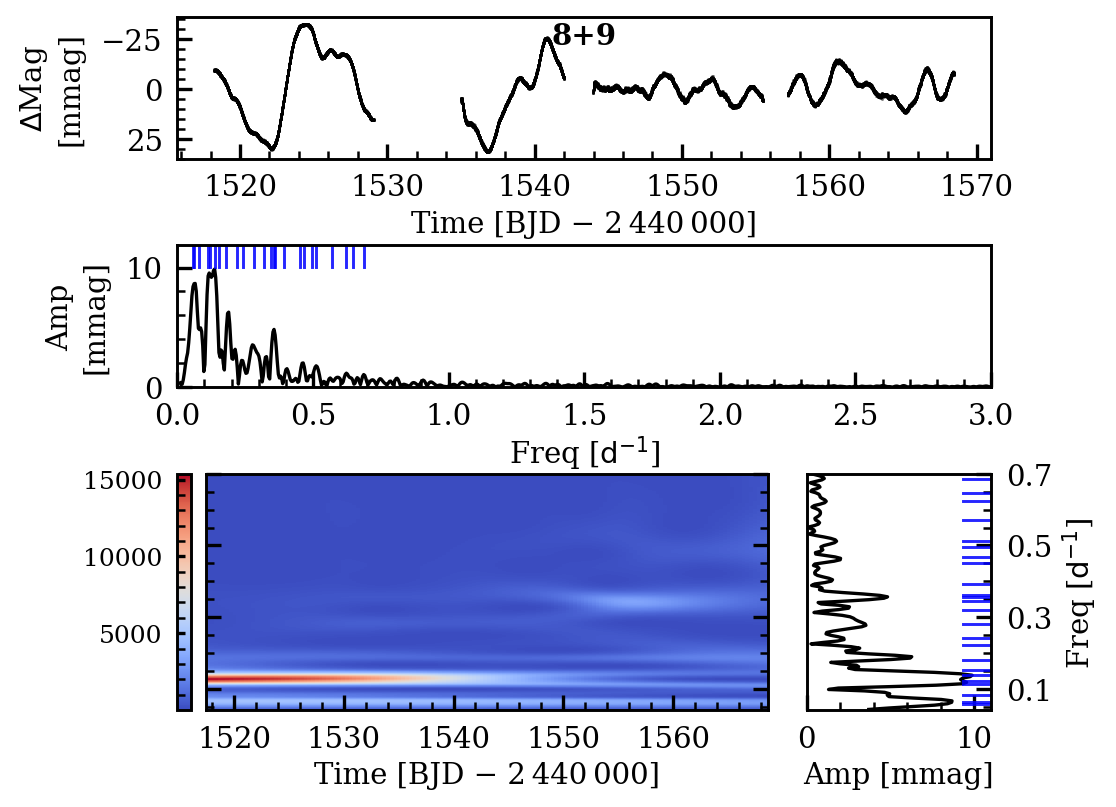} 
    \includegraphics[width=0.33\linewidth]{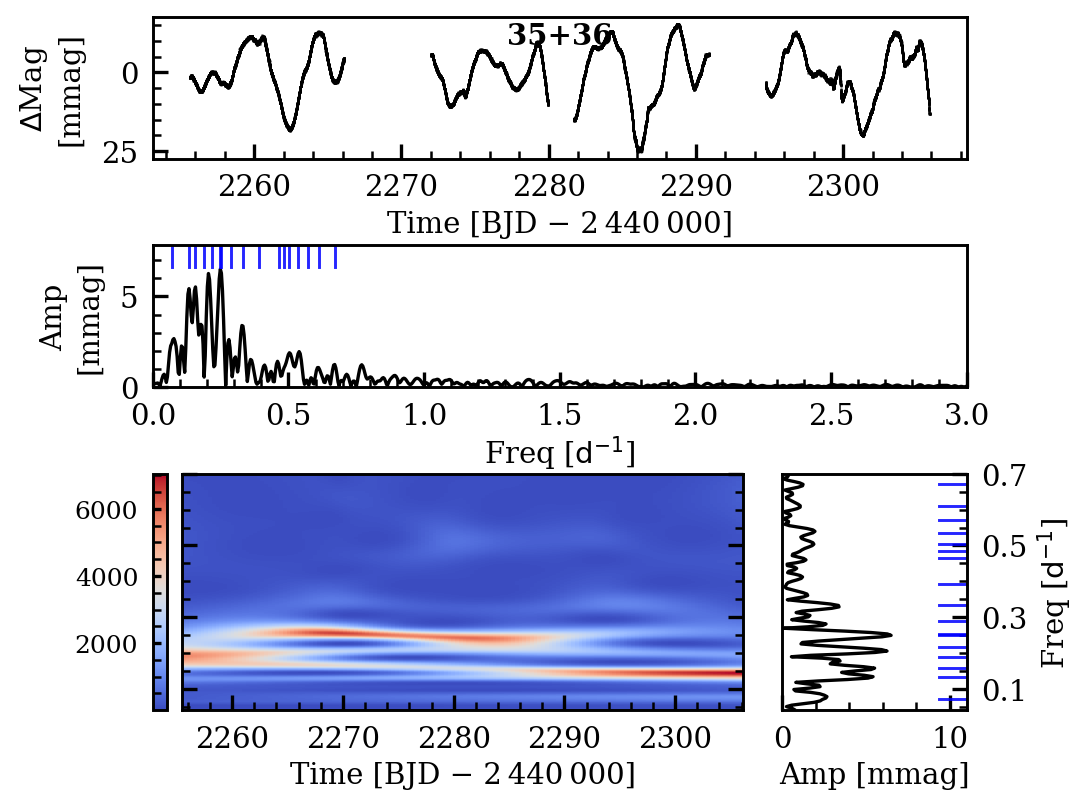}
    \includegraphics[width=0.33\linewidth]{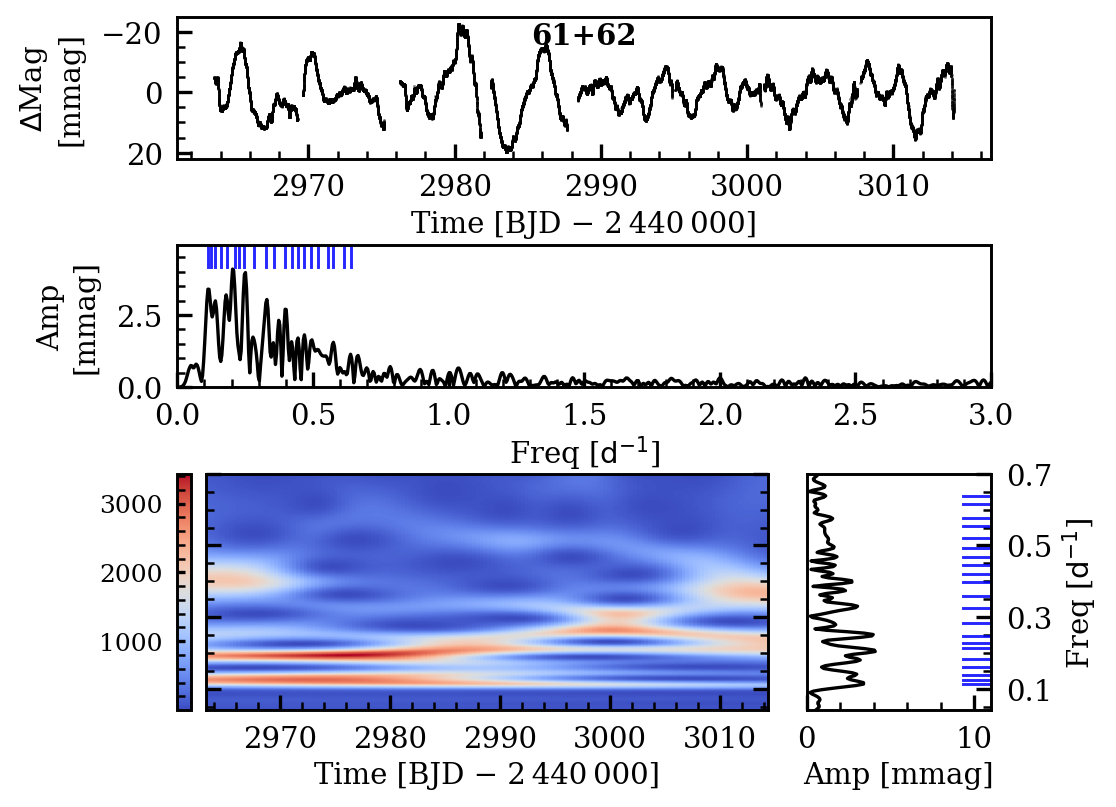}
    \caption{Frequency analysis of \object{HD 74371} from TESS sectors $8+9$, $35+36$, and $61+62$. Panels are as described in Fig.~\ref{fig:freqHD52382}.}
    \label{fig:freqHD74371}
\end{figure*}

\begin{figure*}
    \centering
    \includegraphics[width=0.33\linewidth]{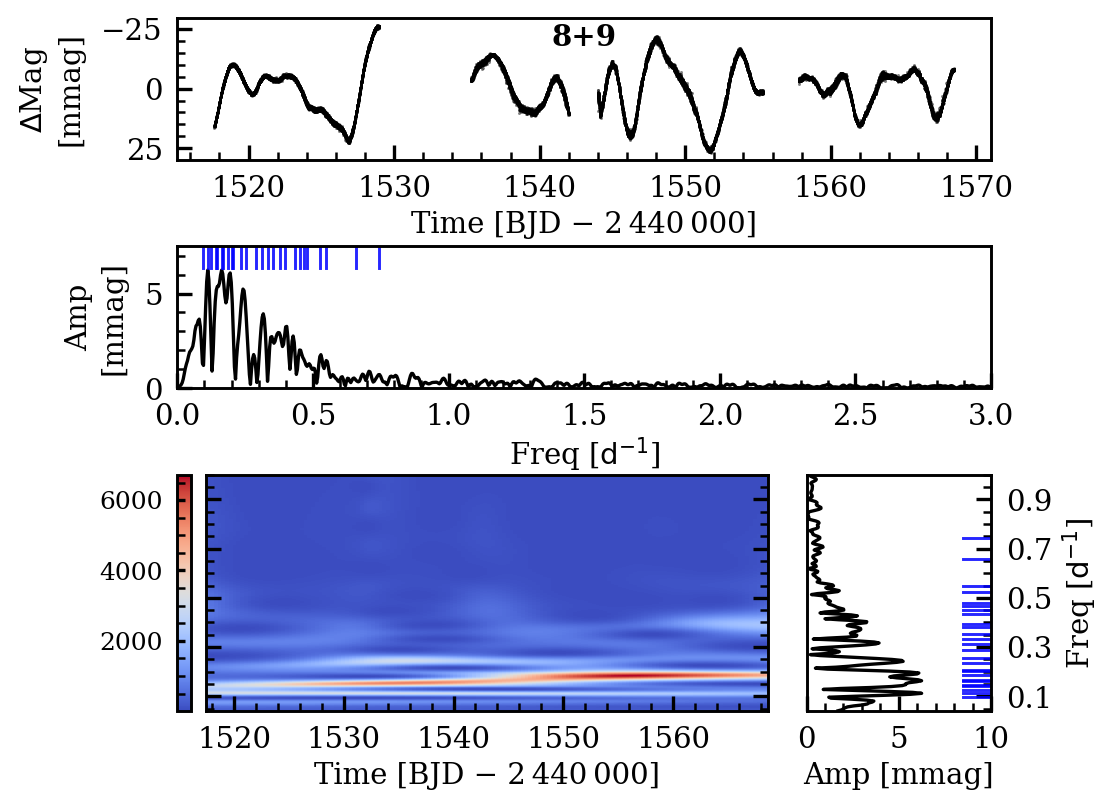} 
    \includegraphics[width=0.33\linewidth]{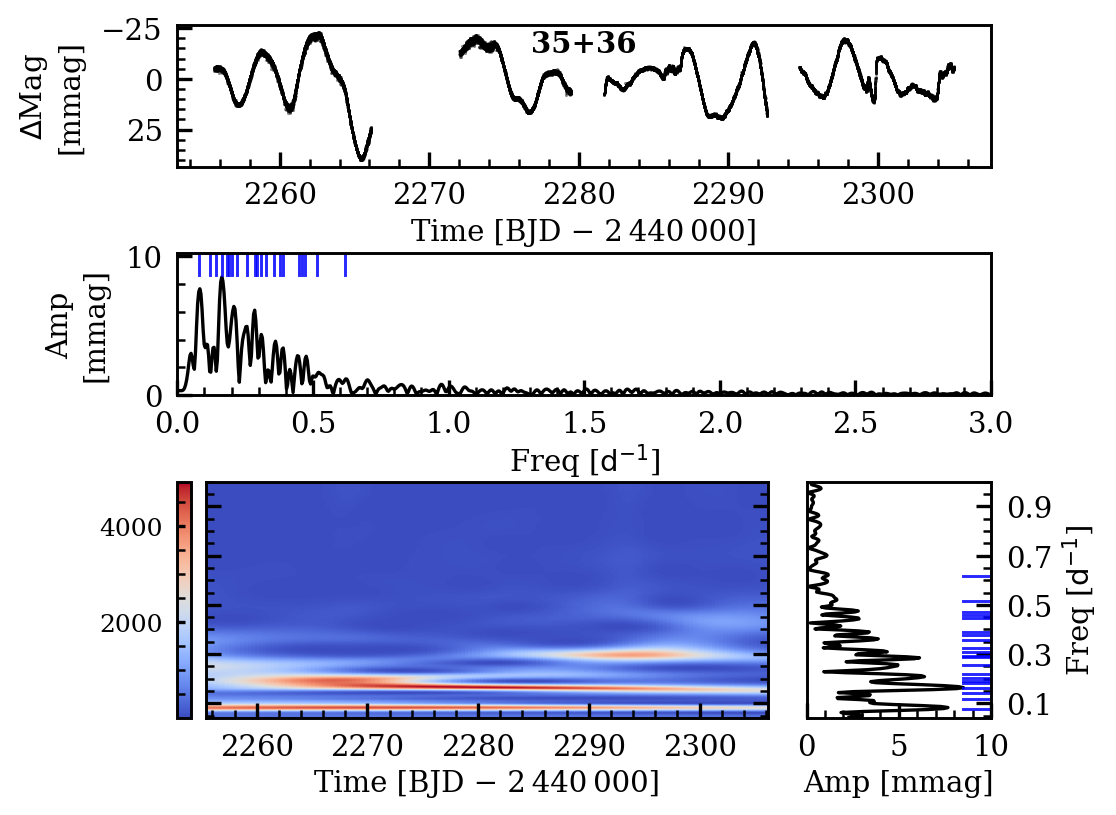}
    \includegraphics[width=0.33\linewidth]{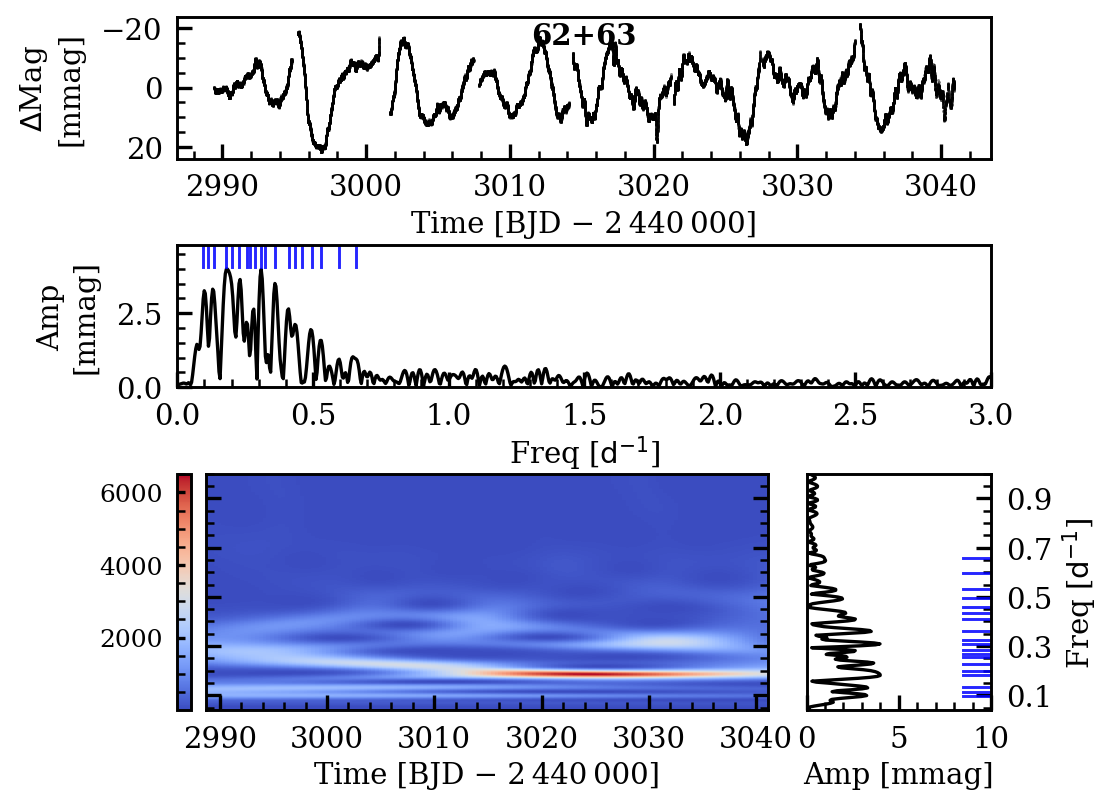}
    \caption{Frequency analysis of \object{HD 79186} from TESS sectors $8+9$, $35+36$, and $62+63$. Panels are as described in Fig.~\ref{fig:freqHD52382}.}
    \label{fig:freqHD79186}
\end{figure*}

\begin{figure*}
    \centering
    \includegraphics[width=0.33\linewidth]{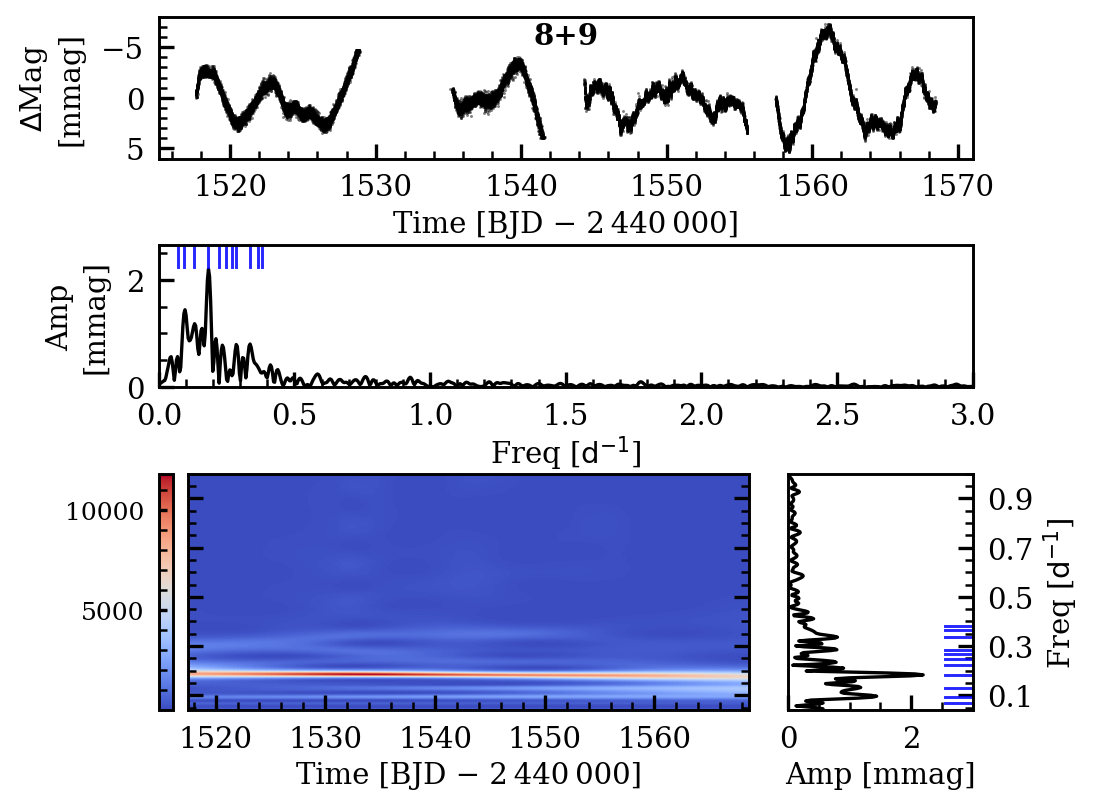} 
    \includegraphics[width=0.33\linewidth]{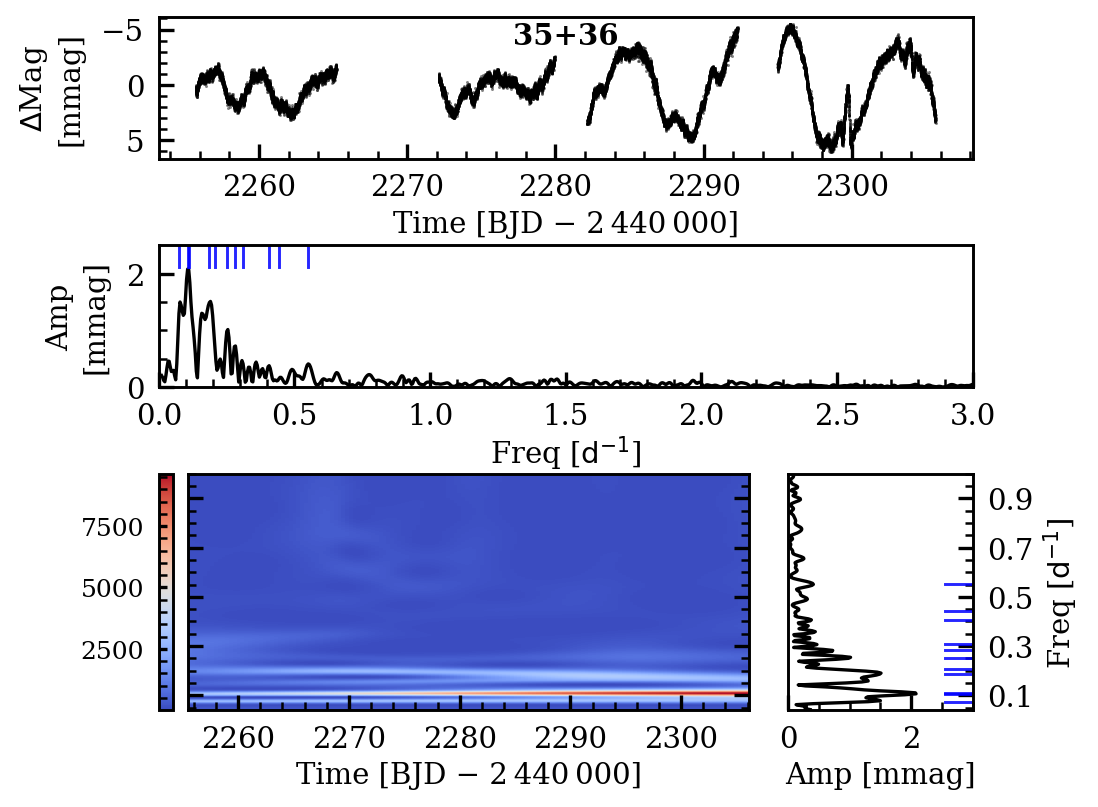}
    \includegraphics[width=0.33\linewidth]{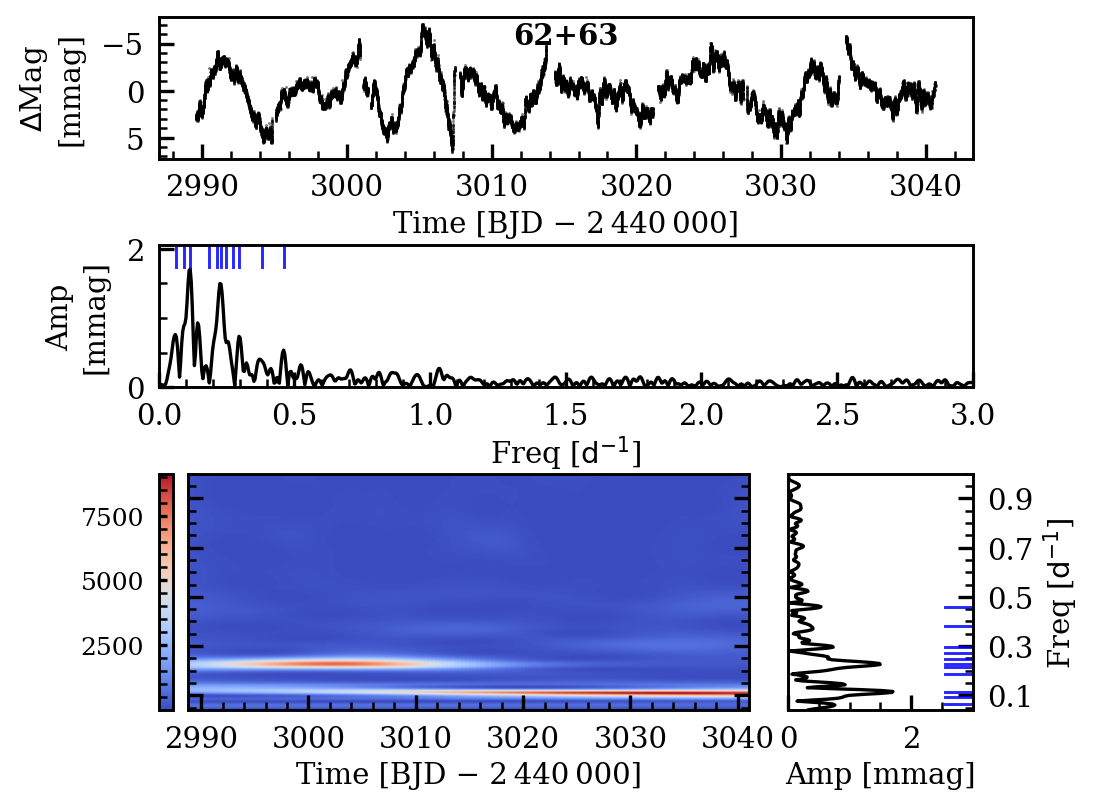}
    \caption{Frequency analysis of \object{HD 80077} from TESS sectors $8+9$, $35+36$, and $62+63$. Panels are as described in Fig.~\ref{fig:freqHD52382}.}
    \label{fig:freqHD80077}
\end{figure*}

\begin{figure*}
    \centering
    \includegraphics[width=0.33\linewidth]{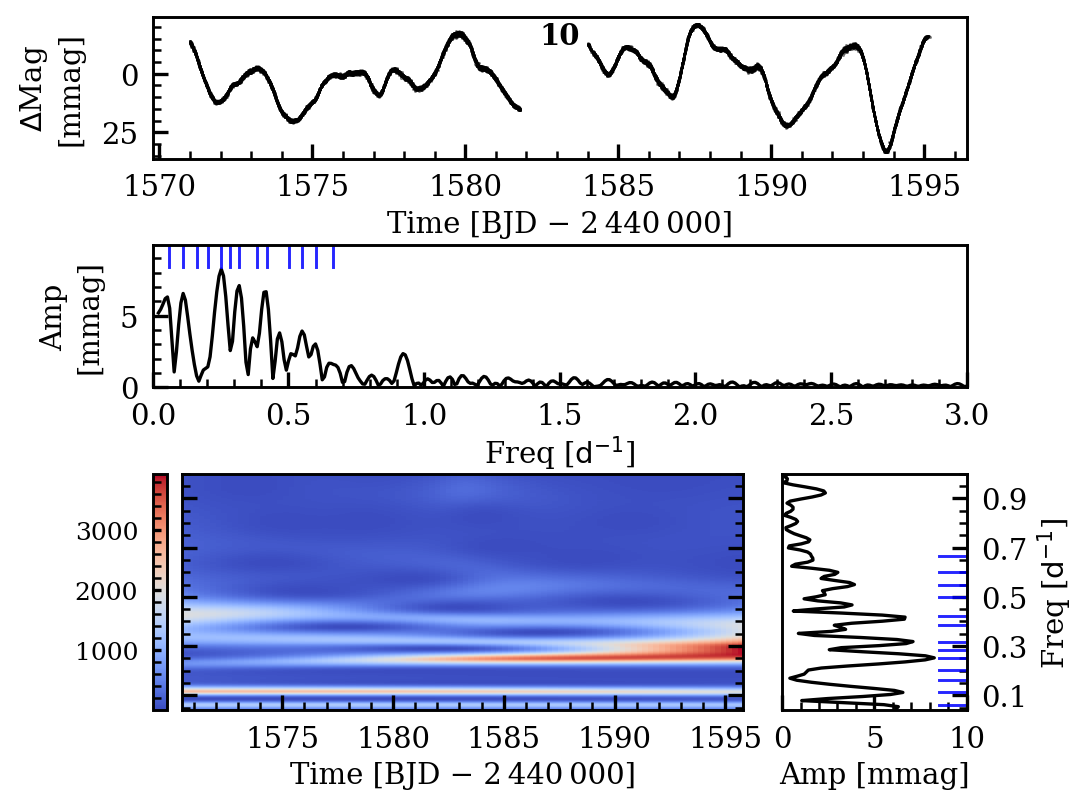} 
    \includegraphics[width=0.33\linewidth]{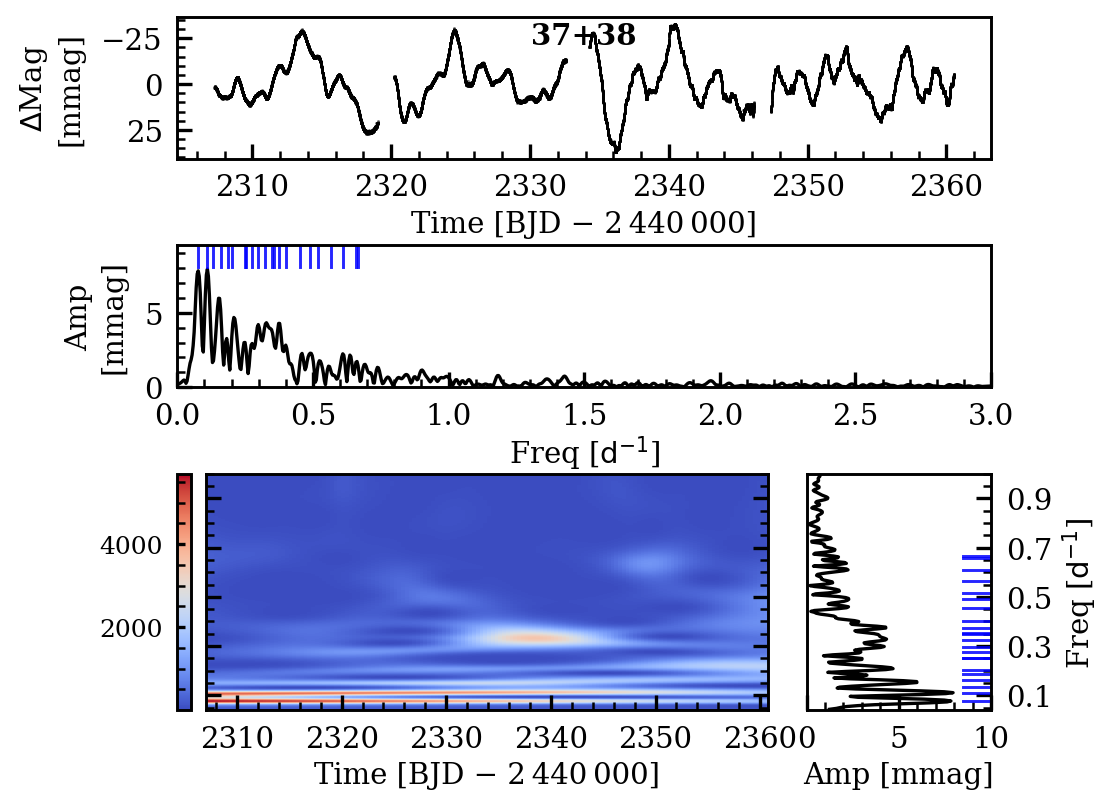}
    \includegraphics[width=0.33\linewidth]{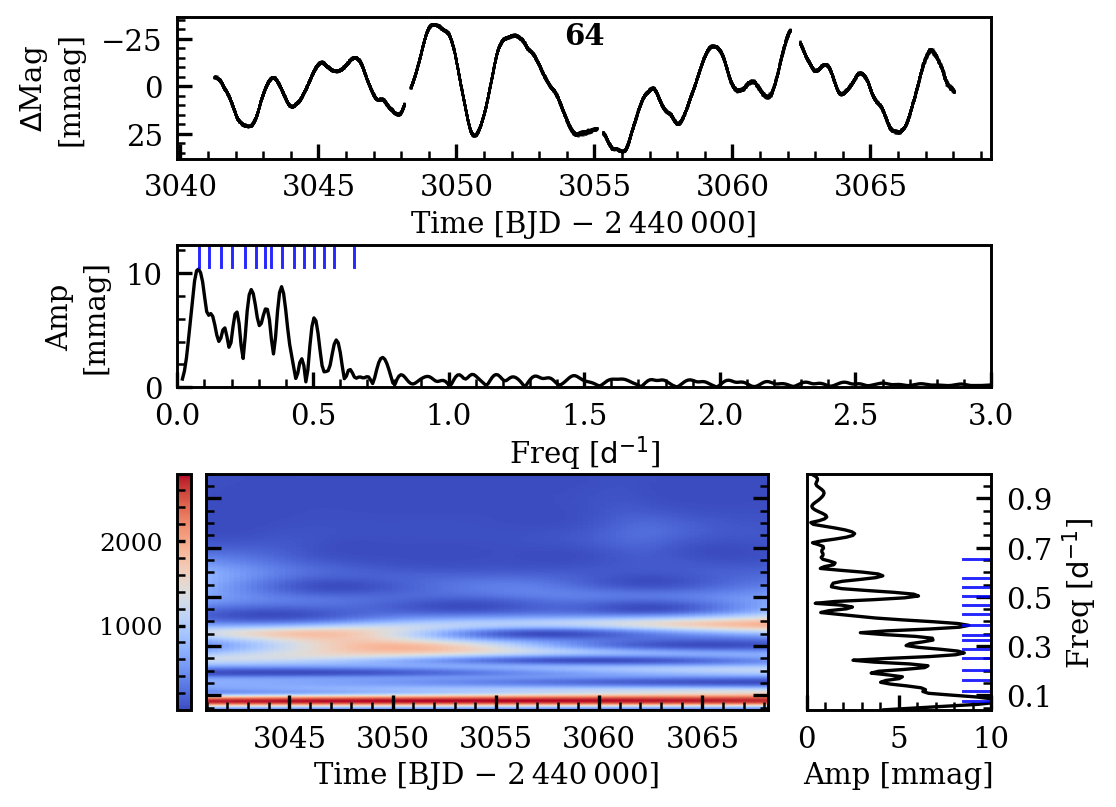}
    \caption{Frequency analysis of \object{HD 99953} from TESS sectors $10$, $37+38$, and $64$. Panels are as described in Fig.~\ref{fig:freqHD52382}.}
    \label{fig:freqHD99953}
\end{figure*}

\begin{figure*}
    \centering
    \includegraphics[width=0.33\linewidth]{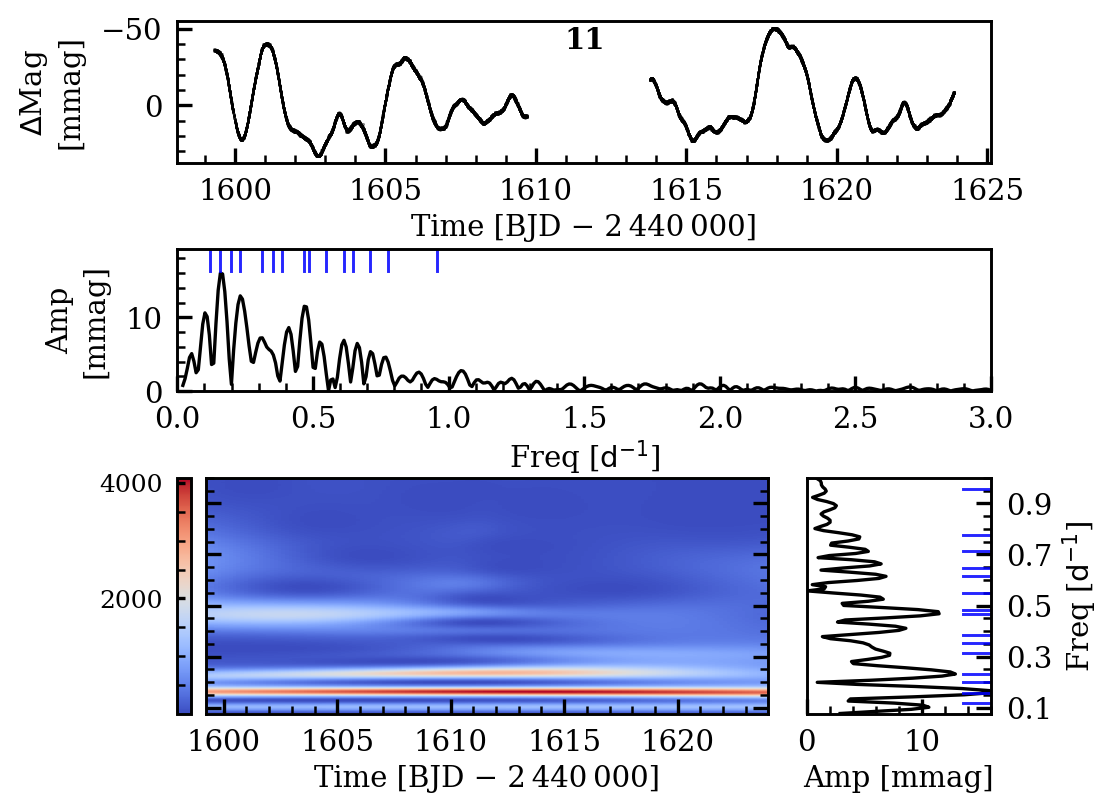} 
    \includegraphics[width=0.33\linewidth]{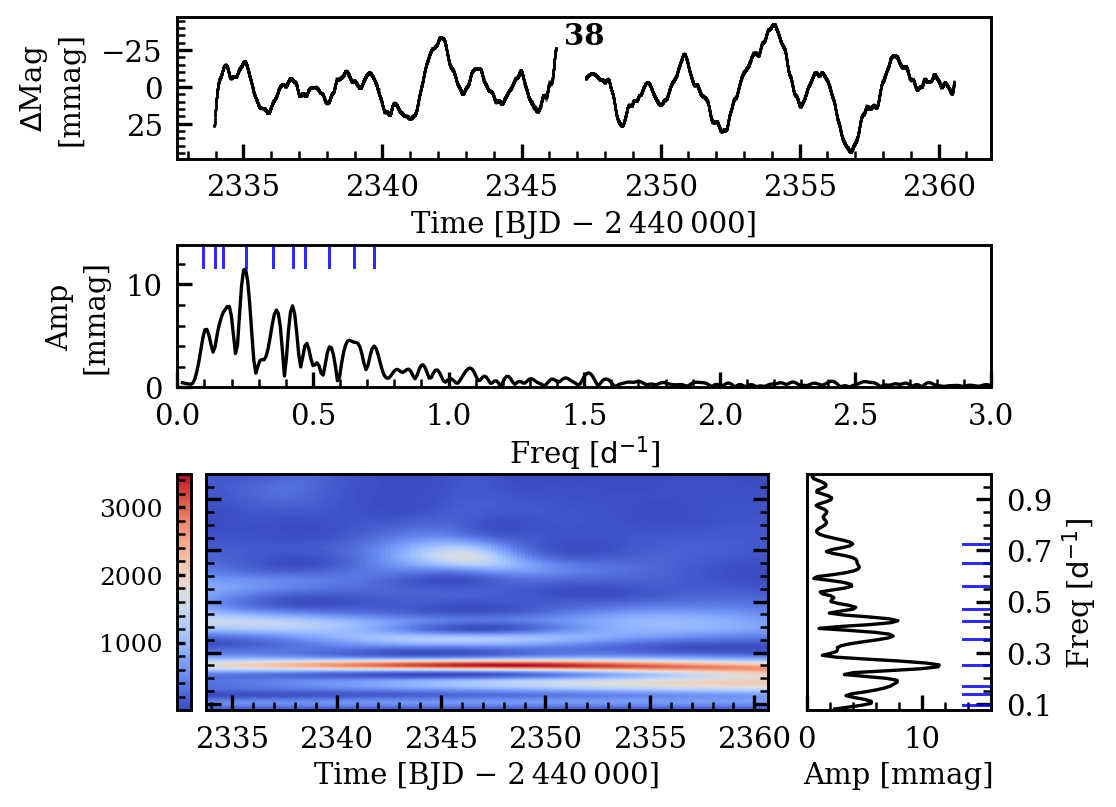}
    \includegraphics[width=0.33\linewidth]{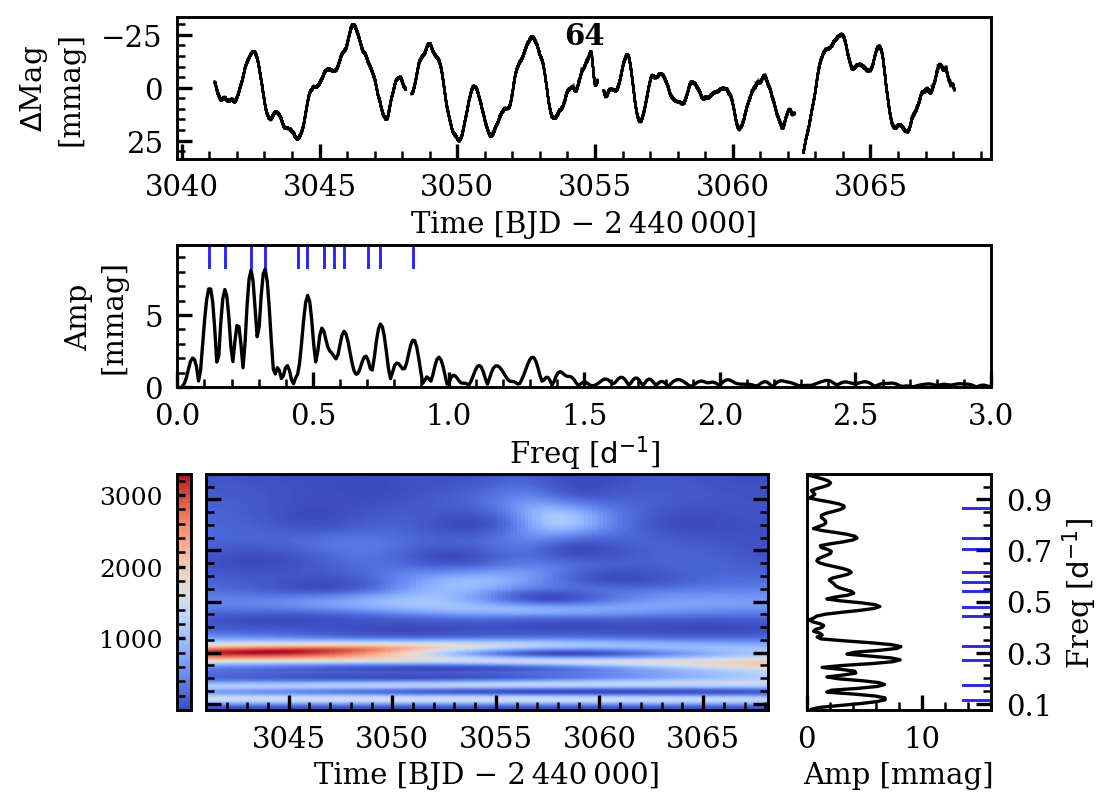}
    \caption{Frequency analysis of \object{HD 115842} from TESS sectors $11$, $38$, and $64$. Panels are as described in Fig.~\ref{fig:freqHD52382}.}    
    \label{fig:freqHD115842}
\end{figure*}

\onecolumn  

\begin{table}[!ht]
\caption{Independent frequencies found for the sample of stars.}
\label{table:all}
\centering
\begin{tabular}{c c c c c c c}
\toprule
\toprule
Star & \multicolumn{3}{c}{TESS sector} & Avg. freq. $\pm\sigma$ & Avg. period $\pm\sigma$ & S/N \\
$f_{id}$ &  &  &  & [$\mathrm{d}^{-1}$] & [$\mathrm{d}$] &   \\
\hline
\rule{0pt}{1em}
HD\,52382 & 7 & 33 & 87 &  &  &   \\
\hline
\rule{0pt}{1em}%
$f_{1}$ & $0.1212(16)$ & $\cdots$ & $0.1093(19)$ & $0.11525 \pm 0.00841$ & $8.676\pm 0.630$ & $[9.2, 7.1]$ \\ 
\rule{0pt}{1em}%
$f_{2}$ & $0.3026(24)$ & $0.2936(23)$ & $0.2821(22)$ & $0.29277 \pm 0.01025$ & $3.415 \pm 0.121$ & $[6.0, 5.7, 6.2]$ \\ 
\rule{0pt}{1em}%
$f_{3}$ & $0.3469(16)$ & $0.3405(28)$ & $0.3418(32)$ & $0.34307 \pm 0.00342$ & $2.915 \pm 0.029$ & $[9.1, 4.7, 4.2]$ \\ 
\rule{0pt}{1em}%
$f_{4}$ & $0.4325(23)$ & $0.4345(18)$ & $0.4292(32)$ & $0.43207 \pm 0.00267$ & $2.314 \pm 0.014$ & $[6.34,7.52,4.27]$ \\ 
\rule{0pt}{1em}%
$f_{5}$ & $0.9149(33)$ & $0.9120(26)$ & $0.9098(29)$ & $0.91223 \pm 0.00255$ & $1.096 \pm 0.003$ & $[4.5, 4.6, 5.1]$ \\ \hline
\rule{0pt}{1em}%
HD\,74371 & 8+9 & 35+36 & 61+62 & & & \\
\hline
\rule{0pt}{1em}%
$f_{1}$ & $0.1135(22)$ & $\cdots$ & $0.1127(16)$ & $0.11310 \pm 0.00057$ & $8.842 \pm 0.045$ & $[4.6,7.4]$ \\ 
\rule{0pt}{1em}%
$f_{2}$ & $0.1374(12)$ & $0.1333(11)$ & $0.1380(17)$ & $0.13623 \pm 0.00255$ & $7.340 \pm 0.138$ & $[18.3,12.3,6.9]$ \\ 
\rule{0pt}{1em}%
$f_{3}$ & $0.1812(18)$ & $0.1890(19)$ & $0.1838(18)$ & $0.18467 \pm 0.00404$ & $5.415 \pm 0.118$ & $[8.6,5.2,6.3]$ \\ 
\rule{0pt}{1em}%
$f_{4}$ & $0.2409(17)$ & $0.2487(12)$ & $0.2471(15)$ & $0.24557 \pm 0.00415$ & $4.072 \pm 0.069$ & $[8.9,14.3,8.4]$ \\ 
\rule{0pt}{1em}%
$f_{5}$ & $0.3205(20)$ & $0.3262(19)$ & $0.3262(17)$ & $0.32430 \pm 0.00329$ & $3.083 \pm 0.031$ & $[5.6,5.7,7.3]$ \\ 
\rule{0pt}{1em}%
$f_{6}$ & $0.3923(16)$ & $0.3953(16)$ & $0.3981(18)$ & $0.39523 \pm 0.00290$ & $2.530 \pm 0.018$ & $[7.4,5.3,6.8]$ \\
\hline
\rule{0pt}{1em}%
HD\,79186 & 8+9 & 35+36 & 62+63 &  &  &    \\
\hline
\rule{0pt}{1em}%
$f_{1}$ & $0.1121(15)$ & $0.1193(23)$ & $0.1147(16)$ & $0.11536 \pm 0.00364$ & $8.669 \pm 0.273$ & $[10.6,6.4,4.3]$ \\ 
\rule{0pt}{1em}%
$f_{2}$ & $0.1633(12)$ & $0.1638(13)$ & $\cdots$ & $0.16353 \pm 0.00034$ & $6.115 \pm 0.013$ & $[13.6,12.6]$ \\ 
\rule{0pt}{1em}%
$f_{3}$ & $0.2026(14)$ & $0.2023(12)$ & $0.2002(13)$ & $0.20170 \pm 0.00133$ & $4.958 \pm 0.033$ & $[9.7,13.3,8.9]$ \\ 
\rule{0pt}{1em}%
$f_{4}$ & $0.2538(15)$ & $0.2568(15)$ & $0.2568(17)$ & $0.25579 \pm 0.00174$ & $3.910 \pm 0.027$ & $[9.5,8.8,6.5]$ \\ 
\rule{0pt}{1em}%
$f_{5}$ & $0.3107(14)$ & $0.3093(15)$ & $0.3090(14)$ & $0.30967 \pm 0.00088$ & $3.229 \pm 0.009$ & $[6.4,9.4,9.8]$ \\
\hline
\rule{0pt}{1em}%
HD\,80077 & 8+9 & 35+36 & 62+63 &  &  &   \\
\hline
\rule{0pt}{1em}%
$f_{1}$ & $0.1219(10)$ & $0.1102(4)$ & $0.1155(5)$ & $0.11584 \pm 0.00588$ & $8.633 \pm 0.444$ & $[11.8,16.23,13.3]$ \\ 
\rule{0pt}{1em}%
$f_{2}$ & $0.1829(3)$ & $0.1863(9)$ & $0.1859(15)$ & $0.18503 \pm 0.00187$ & $5.405 \pm 0.055$ & $[20.4,7.7,4.4]$ \\
\rule{0pt}{1em}%
$f_{3}$ & $0.2459(10)$ & $0.2504(8)$ & $0.2486(11)$ & $0.24830 \pm 0.00225$ & $4.028 \pm 0.036$ & $[6.8,8.8,6.2]$ \\
\rule{0pt}{1em}%
$f_{4}$ & $0.2853(8)$ & $0.2824(13)$ & $0.2956(11)$ & $0.28777 \pm 0.00698$ & $3.475 \pm 0.084$ & $[8.8,5.3,5.9]$ \\
\hline
\rule{0pt}{1em}%
HD\,99953 & 10 & 37+38 & 64 &  &  &   \\
\hline
\rule{0pt}{1em}%
$f_{1}$ & $0.1117(34)$ & $0.1107(12)$ & $0.1176(30)$ & $0.11331 \pm 0.00377$ & $8.825 \pm 0.293$ & $[7.8,11.8,8.7]$ \\ 
\rule{0pt}{1em}%
$f_{2}$ & $0.2513(34)$ & $0.2500(16)$ & $0.2500(35)$ & $0.25040 \pm 0.00076$ & $3.994 \pm 0.012$ & $[8.7,7.3,5.6]$ \\ 
\rule{0pt}{1em}%
$f_{3}$ & $0.3185(37)$ & $0.3246(14)$ & $0.3246(33)$ & $0.32254 \pm 0.00356$ & $3.101 \pm 0.034$ & $[7.5,7.8,7.6]$ \\
\rule{0pt}{1em}%
$f_{4}$ & $0.4077(34)$ & $0.3999(17)$ & $0.3853(31)$ & $0.39763 \pm 0.01144$ & $2.515 \pm 0.072$ & $[8.5,6.4,8.7]$ \\
\hline
\rule{0pt}{1em}%
HD\,115842 & 11 & 38 & 64 &  &  &  \\
\hline
\rule{0pt}{1em}%
$f_{1}$& $0.1184(14)$ & $0.09398(39)$ & $0.1154(13)$ & $0.10926 \pm 0.01332$ & $9.152 \pm 1.116$ & $[5.5,5.0,6.7]$  \\
\rule{0pt}{1em}%
$f_{2}$  & $0.1587(13)$ & $0.16917(34)$ & $0.1750(13)$ & $0.16762 \pm 0.00820$ & $5.966 \pm 0.292$ & $[8.6,6.7,6.6]$   \\
\rule{0pt}{1em}%
$f_{3}$& $0.2319(14)$ & $0.25187(28)$ & $0.2718(13)$ & $0.25186 \pm 0.01995$ & $3.970 \pm 0.315$ & $[5.6,8.7,7.3]$   \\
\rule{0pt}{1em}%
$f_{4}$ & $0.4678(14)$ & $0.4697(14)$ & $0.4782(13)$ & $0.47190 \pm 0.00457$ & $2.119 \pm 0.021$ & $[8.3,7.9,7.8]$  \\
\rule{0pt}{1em}%
$f_{5}$ & $0.7119(10)$ & $0.7153(36)$ & $0.7088(15)$ & $0.71200 \pm 0.00325$ & $1.405 \pm 0.006$ & $[6.5, 6.3, 6.2]$  \\
\bottomrule
\end{tabular}
\tablefoot{
The first column lists the name of the star and frequency ID ($f_{id}$), ordered from lowest to highest. The listed values were obtained with the Lomb-Scargle periodogram for each TESS sector; uncertainties are given in parentheses for the last significant digits. The dispersion $\sigma$ is indicated for the average frequencies and periods. The $\mathrm{S/N}$ ratio corresponding to each sector is given in the last column.
}
\end{table}

\end{document}